\documentclass[%
 reprint,
superscriptaddress,
nofootinbib,
 amsmath,amssymb,
 onecolumn,
prd,
]{revtex4-2}

\usepackage{graphicx}
\usepackage{dcolumn}
\usepackage{bm}
\usepackage[colorlinks]{hyperref}

\usepackage{color}
\usepackage[caption=false]{subfig}

\renewcommand{\vec}[1]{\boldsymbol{#1}}

\newcommand{\Mpch}{h^{-1}\mathrm{Mpc}}

\newcommand{\delD}[1]{(2\pi)^3\delta_\mathrm{D}\left({#1}\right)}

\newcommand{\av}[1]{\left\langle{#1}\right\rangle}

\newcommand{\vk}{\vec k}
\newcommand{\vK}{\vec K}

\newcommand{\vs}{\vec s}

\newcommand{\vx}{\vec x}

\renewcommand{\vr}{\vec r}

\renewcommand{\L}{\Lambda}
\renewcommand{\P}{\mathcal{P}}

\newcommand{\hr}{\hat{\vec r}}
\newcommand{\hn}{\hat{\vec n}}
\newcommand{\hk}{\hat{\vec k}}
\newcommand{\hs}{\hat{\vec s}}
\newcommand{\tjo}[3]{\begin{pmatrix} {#1} & {#2} & {#3}\\ 0 & 0 & 0\end{pmatrix}}
\newcommand{\tj}[6]{\begin{pmatrix} {#1} & {#2} & {#3}\\ {#4} & {#5} & {#6}\end{pmatrix}}

\newcommand{\expect}[1]{\mathbb{E}\left[{#1}\right]}

\usepackage{color}
\def\beq{\begin{eqnarray}}
\def\eeq{\end{eqnarray}}

\definecolor{darkgreen}{RGB}{0,120,0}
\definecolor{darkred}{RGB}{150,30,30}
\newcommand{\unblind}[1]{#1}
\newcommand{\resub}[1]{#1}

\begin{document}


\title{
Probing Parity-Violation with the Four-Point Correlation Function of BOSS Galaxies}

\author{Oliver H.\,E. Philcox}
\email{ohep2@cantab.ac.uk}
\affiliation{Department of Astrophysical Sciences, Princeton University,\\ Princeton, NJ 08540, USA}%
\affiliation{School of Natural Sciences, Institute for Advanced Study, 1 Einstein Drive,\\ Princeton, NJ 08540, USA}



\begin{abstract}
Parity-violating physics in the early Universe can leave detectable traces in late-time observables. Whilst vector- and tensor-type parity-violation can be observed in the $B$-modes of the cosmic microwave background, scalar-type signatures are visible only in the four-point correlation function (4PCF) and beyond. This work presents a blind test for parity-violation in the 4PCF of the BOSS CMASS sample, considering galaxy separations in the range $[20,160]\Mpch$. The parity-odd 4PCF contains no contributions from standard $\Lambda$CDM physics and can be efficiently measured using recently developed estimators. Data are analyzed using both a non-parametric rank test (comparing the BOSS 4PCFs to those of realistic simulations) and a compressed $\chi^2$ analysis, with the former avoiding the assumption of a Gaussian likelihood. These find similar results, with the rank test giving a detection probability of $99.6\%$ ($2.9\sigma$). This provides significant evidence for parity-violation, either from cosmological sources or systematics. We perform a number of systematic tests: although these do not reveal any observational artefacts, we cannot exclude the possibility that our detection is caused by the simulations not faithfully representing the statistical properties of the BOSS data. Our measurements can be used to constrain physical models of parity-violation. As an example, we consider a coupling between the inflaton and a $U(1)$ gauge field and place bounds on the latter's energy density, which are several orders of magnitude stronger than those previously reported. Upcoming probes such as DESI and Euclid will reveal whether our detection of parity-violation is due to new physics, and strengthen the bounds on a variety of models.
\end{abstract}

\maketitle


\section{Introduction}\label{sec: intro}


A detection of parity-violation in cosmological observables would be a smoking gun for physics beyond the standard model, and could provide crucial insights into the nature of dark matter, dark energy, and inflation. In the conventional paradigm, all cosmological correlators are symmetric under the parity operator $\mathbb{P}$, since gravity (along with all other standard model interactions except the weak force \citep{PhysRev.104.254}), is $\mathbb{P}$-invariant. Despite this, a number of theoretical arguments suggest that parity-violating interactions \textit{should} occur in the early Universe, most notably to source baryogenesis. Creation of the current baryon asymmetry requires a process which violates charge and parity conservation \citep{1967JETPL...5...24S,1998PhRvL..81.3067C}; a possible route is via leptogenesis, which, if sourced by gravity, must be parity-violating \citep[e.g.,][]{2004PhRvL..93t1301D,2016IJMPD..2540013A,2008PhLB..660..444A,2020JHEP...04..189L,2006PhRvL..96h1301A}. 

Additional sources of parity-violation include inflationary interactions between multiple fields, such as via the Chern-Simons term \citep[e.g.,][]{1999PhRvL..83.1506L,2017JCAP...07..034B,2009PhR...480....1A,2015JCAP...01..027B,2015JCAP...07..039B,2020JCAP...07..014B}, generation of primordial magnetic fields \citep[e.g.,][]{2001PhR...348..163G,2016A&A...594A..19P,2012JCAP...06..015S}, vector perturbations generated by cosmic strings or defects \citep[e.g.,][]{2008PhRvD..77h3509P,2015JCAP...03..008R,2021PhRvD.104b3507R}, reheating \citep[e.g.,][]{2012PhRvD..85b3534C,2015JCAP...12..034A}, Chern-Simons modified general relativity \citep{2009PhR...480....1A}, and inflationary particle exchange \citep{2012PhRvL.108y1301J,2015arXiv150308043A}, all of which leave distinctive imprints on late-time observables \citep[e.g.,][]{1999PhRvL..83.1506L}. Potential evidence for such models was recently provided by \citep{2020PhRvL.125v1301M}, which found a \resub{$2.4\sigma$ hint (updated to $3.6\sigma$ in \citep{Eskilt:2022cff}) of parity-violation in the cosmic microwave background (hereafter CMB). Whilst some argue that this effect may be caused by interstellar dust emission \citep{2021ApJ...919...53C} (though see \citep{Eskilt:2022wav,Diego-Palazuelos:2022dsq})}, it has nevertheless provided a resurgence of interest in these theories. 

To constrain such phenomena, we require observables that are parity-sensitive. Common choices are vector and tensor quantities, such as $B$ modes of the CMB \citep[e.g.,][]{1997PhRvL..78.2054S}, or those of galaxy ellipticities \citep[e.g.,][]{2002ApJ...568...20C}. These satisfy $\mathbb{P}[B] = -B$, and can be combined in two-point correlators (e.g., $TB$ and $EB$ for the CMB, or $EB$ for weak lensing). Barring contamination by systematics, the observables should have no contribution from standard $\L$CDM physics, but can be sourced by effects such as birefringence (whereupon the plane of the photon polarization is coherently rotated between the surface of last scattering and the observer, as in \citep{2020PhRvL.125v1301M}), gravitational wave chirality \citep[e.g.,][]{1999PhRvL..83.1506L,1990PhRvD..41.1231C,2010PhRvD..81l3529G,2018PhRvD..97b3532C,2017PhRvL.118v1301M}, and multi-field inflation \citep{2012PhRvL.108y1301J,2015JCAP...10..032S,2020JHEP...04..189L}. Information is not limited to the two-point function however; higher-order correlators such as $TTB$ can give additional constraining power on effects such as birefringence \citep{2011PhRvD..83b7301K}.

\begin{figure}
    \centering
    \includegraphics[width=0.5\textwidth]{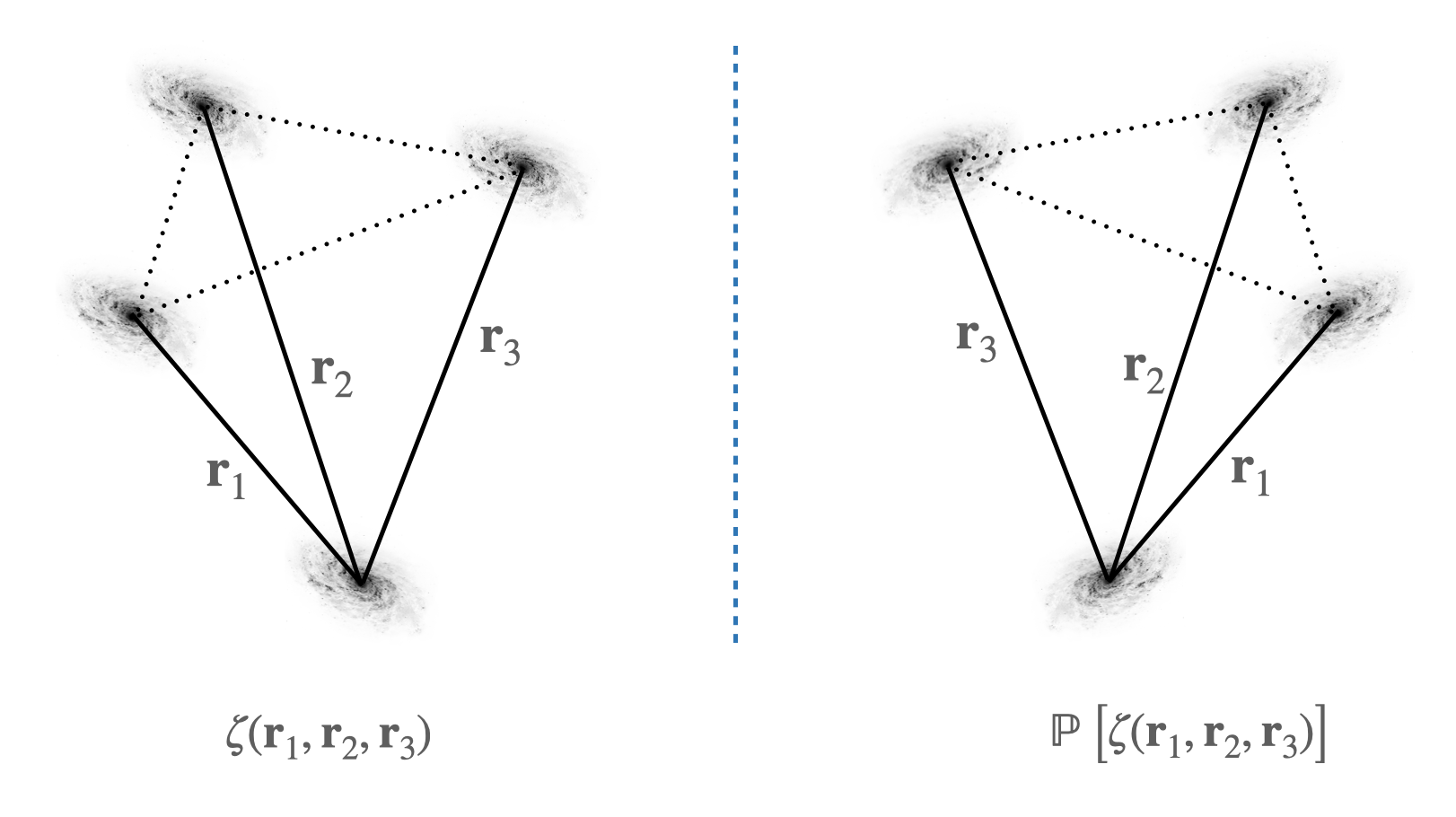}
    \caption{Cartoon of the galaxy four-point correlation functions (4PCFs) considered in this work. In the left panel, we show the 4PCF, $\zeta(\vr_1,\vr_2,\vr_3)$, which depends on the separation vectors of three secondary galaxies from a given primary. The right panel shows the parity-inverted 4PCF, $\mathbb{P}\left[\zeta(\vr_1,\vr_2,\vr_3)\right]$, which corresponds to replacing $\vr_i$ with $-\vr_i$. Unlike for the 2PCF and 3PCF, the two configurations cannot be related by a rotation. The parity-even 4PCF is a sum of the two geometries (which have the same side-lengths and relative angles), whilst the parity-odd 4PCF is a difference. In this work, the 4PCF is given as a function of three lengths ($r_1$, $r_2$, and $r_3$) and three internal angles (fixing the angle of the $\vr_i$ vectors with the respect to the primary galaxy). The latter are represented by their harmonic-space momenta, $\ell_1$, $\ell_2$ and $\ell_3$, with odd-parity 4PCFs corresponding to odd $\ell_1+\ell_2+\ell_3$. Assuming standard $\Lambda$CDM physics, the two correlators shown in the figure should be equivalent, thus the expectation value of the parity-odd 4PCF is zero.}
    \label{fig: 4pcf-cartoon}
\end{figure}

When constructing observables from scalar fields (such as the galaxy density or CMB temperature), obtaining a parity-sensitive quantity is more difficult. As an example, the isotropic galaxy two-point correlation function (hereafter 2PCF) is insensitive to parity, since the action of $\mathbb{P}$ is equivalent to a rotation, under which the statistic is invariant. In three-dimensions, the isotropic $N$-point correlation functions (NPCFs) are parity-sensitive only if $N>3$; this applies also to the CMB, since the intrinsic fluctuations are the projection of a three-dimensional quantity. The simplest statistic with which to probe scalar parity-violation is thus the 4PCF, as pointed out in \citep{2012PhRvL.108y1301J,2016PhRvD..94h3503S,2021arXiv211012004C}. A cartoon of this is shown in Fig.\,\ref{fig: 4pcf-cartoon}.\footnote{Large scale structure correlators are sensitive also to redshift-space distortions \citep{1987MNRAS.227....1K,1998ASSL..231..185H}, giving dependence of the statistic on line-of-sight velocities \citep{2018JCAP...02..028B}. This enables vector-type parity-violation to be probed in the 3PCF \citep{2019arXiv190605198J}, though it requires careful modelling of galaxy velocities.}

Whilst a number of works have considered the 4PCF of the CMB \citep[e.g.,][]{2011MNRAS.412.1993M,2015arXiv150200635S} including its parity-odd contributions \citep{2013PhRvD..87j3006D,2016PhRvD..94h3503S} (though only theoretically), the large scale structure (LSS) equivalent has been rarely explored. Given the influx of spectroscopic data expected in the next decade from DESI \citep{2016arXiv161100036D}, Euclid \citep{2011arXiv1110.3193L}, and Rubin \citep{collaboration2009lsst}, galaxy surveys seem to be a natural arena in which to hunt for parity-violating interactions, allowing constraints to exceed the CMB cosmic variance limit. Historically, use of the higher-point galaxy correlation functions has been hampered by the computational resources required for their estimation; na\"ively, the 4PCF requires $\mathcal{O}(N_{\rm g}^4)$ operations to compute from $N_{\rm g}$ galaxies. Recent works have significantly improved upon this \citep{2019ApJS..242...29S,npcf_algo}, with the algorithm of \citep{npcf_algo} requiring only $\mathcal{O}(N_{\rm g}^2)$ operations. This allows the 4PCF of current galaxy surveys to be computed in $\sim$\,$30$ CPU-hours. The approach proceeds by first projecting the correlation function into a suitable angular basis \citep{2020arXiv201014418C}; thence, the integrals decouple and the 4PCF may be computed by summing over pairs of galaxies. This naturally generalizes to higher-dimensions, as well as to anisotropic correlators \citep{gen_npcf}. 
As first pointed out in \citep{2020arXiv201014418C} there is a natural separation of the parity-even and parity-odd isotropic basis functions. The even-parity component 
can be used to place constraints on gravitational non-Gaussianity from a hitherto unexplored statistic \citep{4pcf_boss}.
The use of the parity-odd basis to measure parity violation in the galaxy four-point correlation was first proposed in \citep{2021arXiv211012004C} and is carried out in this work (see also \citep{2012PhRvL.108y1301J}).

There are two main ways in which parity-violation can be probed using the galaxy 4PCF. Firstly, one may place constraints on the amplitudes of specific physical models given their associated theoretical predictions. This is an approach oft-used in the analysis of CMB 3- and 4-point functions, for example in non-Gaussianity studies, which typically exploit separability of the underlying theoretical templates for significant computational gain \citep[e.g.,][]{2010AdAst2010E..73L}. This approach was also suggested in \citep{2012PhRvL.108y1301J,2013PhRvD..87j3006D,2014JCAP...12..050D}, and allows for targeted constraints on specific models of early-Universe particle exchange, via a search for their specific isotropy-violating signatures. An alternative method would be to first measure the \textit{full} galaxy 4PCF in some set of bins, then perform a blind test, looking for the signatures of \textit{any} physical model (and systematic effects). This approach is possible since the parity-odd 4PCF receives no contribution in $\L$CDM, including from general relativistic and baryonic effects. Given the multitude of possible models for parity-violation, we will principally adopt the second strategy in this work, though we demonstrate also the first, by placing constraints on a specific model involving Chern-Simons terms in the inflationary Lagrangian. Analysis using the galaxy 4PCF comes with its complexities, however. In particular, the high-dimensionality of the statistic prohibits conventional mock-based $\chi^2$-analyses. To alleviate this, we include a data-compression step, facilitated using a theoretical model of the 4PCF covariance \citep{npcf_cov}, which dramatically reduces the number of bins without introducing bias. It is not guaranteed that the 4PCF likelihood be Gaussian however (see Appendix \ref{appen: NG-likelihood} and \citep{2012PhRvD..86f3009S,2019MNRAS.485.2956H}); to provide a fully robust, yet conservative, test for parity-violation, we make use of a likelihood-free inference technique, involving a suite of realistic simulations. We caution that such blind tests are naturally subject to systematic uncertainties, some of which will be explored in this work. The results below represent the first constraints on scalar parity-violation from LSS data.

\vskip 4pt

The remainder of this work is structured as follows. In \S\ref{sec: estimator}, we present the parity-odd 4PCF estimator, including the corrections necessary to account for non-uniform survey geometry, before we discuss the data and covariance matrices in \S\ref{sec: data}. Analysis methods are considered in \S\ref{sec: analysis-methods}, with the corresponding constraints on parity-violation presented in \S\ref{sec: results}. In \S\ref{sec: systematics}, we include a number of systematic checks and a brief discussion of potential biases in the approach. \S\ref{sec: inflationary-model} discusses parity-breaking phenomena including the presentation of an inflationary model for the parity-odd 4PCF, based on a Chern-Simons coupling, whose amplitude is then bounded using the BOSS data. We conclude in \S\ref{sec: summary}, with Appendices \ref{appen: NG-likelihood} and \ref{appen: chern-simons} discussing the impacts of likelihood non-Gaussianity and sketching the derivation of the Chern-Simons 4PCF template. \textsc{Jupyter} notebooks containing our analysis pipeline can be found on GitHub.\footnote{Available at \href{https://github.com/oliverphilcox/Parity-Odd-4PCF}{github.com/oliverphilcox/Parity-Odd-4PCF}.} 

\vskip 4pt 

\textit{Note on Blinding}: To limit confirmation bias, the BOSS data were sent to an external collaborator (M.\,K\"onig) after computation, and not revealed until the analysis pipeline was constructed and tested. The initial draft of the paper was also written before unblinding (encompassing all sections except \S\ref{sec: systematics} and Appendix\,\ref{appen: NG-likelihood}), with the BOSS data replaced by that from a single mock dataset. 

\section{Measuring The Parity-Odd 4PCF}\label{sec: estimator}
We begin by outlining our estimator for the parity-odd 4PCF, which is implemented in the public \textsc{encore} code.\footnote{Available at \href{https://github.com/oliverphilcox/encore}{github.com/oliverphilcox/encore}.} Further details of the formalism can be found in \citep{npcf_algo} (for the general NPCF estimator and \textsc{encore}), \citep{2020arXiv201014418C} (for the basis functions), \citep{4pcf_boss} (for the parity-even 4PCF), \citep{2021arXiv211012004C} (for an overview of the parity-odd 4PCF) and \citep{gen_npcf} (for extensions beyond flat 3D space).

\subsection{Isotropic Basis Functions}\label{subsec: basis}
Given a (scalar) density field $\delta(\vr)$, the 4PCF is defined as
\beq\label{eq: 4pcf-def}
    \boxed{\zeta(\vr_1,\vr_2,\vr_3) \equiv \av{\delta(\vs)\delta(\vs+\vr_1)\delta(\vs+\vr_2)\delta(\vs+\vr_3)},}
\eeq
where $\av{\cdots}$ represents a statistical average over realizations of $\delta$. A cartoon of this parametrization is shown in Fig.\,\ref{fig: 4pcf-cartoon}. By statistical homogeneity, the 4PCF is independent of the absolute coordinate $\vs$. 

As demonstrated in \citep{npcf_algo,gen_npcf}, a complete angular basis for the isotropic $N$-point correlation functions is given by the isotropic basis functions of $(N-1)$ coordinates defined in \citep{2020arXiv201014418C} (see also the TriPoSH formalism; \citep{1988qtam.book.....V}).\footnote{The approach naturally extends to \textit{anisotropic} correlators \citep{gen_npcf}, though we do not consider them in this work.} For $N=4$, the basis functions are
\beq\label{eq: basis-def}
    \P_{\ell_1\ell_2\ell_3}(\hr_1,\hr_2,\hr_3) &\equiv& (-1)^{\ell_1+\ell_2+\ell_3}\sum_{m_1m_2m_3}\tj{\ell_1}{\ell_2}{\ell_3}{m_1}{m_2}{m_3}Y_{\ell_1m_1}(\hr_1)Y_{\ell_2m_2}(\hr_2)Y_{\ell_3m_3}(\hr_3),
\eeq
where $Y_{\ell m}(\hr)$ is a spherical harmonic, the $3\times 2$ matrix is a Wigner 3-$j$ symbol, and the $m_i$ summations run over integer $m_i\in [-\ell_i,\ell_i]$. Such functions arise from the theory of angular momentum addition, and are specified by three non-negative integers $\{\ell_1,\ell_2,\ell_3\}$, encoding the relative orientation of $\hr_1, \hr_2$, and $\hr_3$. Due to the 3-$j$ symbol, the integers must obey the triangle condition $|\ell_1-\ell_2|\leq \ell_3\leq \ell_1+\ell_2$, and we additionally enforce $\ell_i\leq \ell_{\rm max}$. In practice, we restrict to relatively low $\ell_{\rm max}$, which gives an angular resolution of $\theta_{\rm min} \approx 2\pi/\ell_{\rm max}$ for the internal angles of the 4PCF tetrahedron. The basis functions have the following properties under parity and conjugation transformations (for parity operator $\mathbb{P}$):
\beq\label{eq: basis-conj-parity}
    \mathbb{P}\left[\P_{\ell_1\ell_2\ell_3}(\hr_1,\hr_2,\hr_3)\right] &=& (-1)^{\ell_1+\ell_2+\ell_3}\P_{\ell_1\ell_2\ell_3}(\hr_1,\hr_2,\hr_3), \qquad \P^*_{\ell_1\ell_2\ell_3}(\hr_1,\hr_2,\hr_3) = (-1)^{\ell_1+\ell_2+\ell_3}\P_{\ell_1\ell_2\ell_3}(\hr_1,\hr_2,\hr_3),
\eeq
implying that the basis is parity-odd and pure imaginary if $\ell_1+\ell_2+\ell_3$ is odd, and parity-even and real else. Furthermore, \eqref{eq: basis-def} is invariant under joint rotations of all three separation vectors, \textit{i.e.}\ $\{\vr_1,\vr_2,\vr_3\}\to \{R\vr_1,R\vr_2,R\vr_3\}$, for arbitrary rotation matrix $R$.

The isotropic part of the galaxy 4PCF can be decomposed into the basis of \eqref{eq: basis-def}:
\beq\label{eq: 4pcf-decomposition}
    \zeta_{\rm iso}(\vr_1,\vr_2,\vr_3) = \sum_{\ell_1\ell_2\ell_3}\zeta_{\ell_1\ell_2\ell_3}(r_1,r_2,r_3)\P_{\ell_1\ell_2\ell_3}(\hr_1,\hr_2,\hr_3),
\eeq
where the coefficients $\zeta_{\ell_1\ell_2\ell_3}$ (hereafter denoted `multiplets') can be obtained via the orthonormality of $\P_{\ell_1\ell_2\ell_3}$.\footnote{Since the anisotropic basis functions are orthogonal to those of \eqref{eq: basis-def}, the decomposition in \eqref{eq: 4pcf-decomposition} holds regardless of whether the full statistic is isotropic.}  Given the transformation properties of \eqref{eq: basis-conj-parity}, we find a natural split of $\zeta_{\rm iso}$ into parity-even and parity-odd parts:
\beq
    \zeta_+(\vr_1,\vr_2,\vr_3) = \sum_{\ell_1+\ell_2+\ell_3 = \text{even}}\zeta_{\ell_1\ell_2\ell_3}(r_1,r_2,r_3)\P_{\ell_1\ell_2\ell_3}(\hr_1,\hr_2,\hr_3),\\\nonumber
    \zeta_-(\vr_1,\vr_2,\vr_3) = \sum_{\ell_1+\ell_2+\ell_3 = \text{odd}}\zeta_{\ell_1\ell_2\ell_3}(r_1,r_2,r_3)\P_{\ell_1\ell_2\ell_3}(\hr_1,\hr_2,\hr_3).
\eeq
These satisfy $\mathbb{P}\left[\zeta_{\pm}(\vr_1,\vr_2,\vr_3)\right] = \pm\, \zeta_{\pm}(\vr_1,\vr_2,\vr_3)$, and may be related to the sum and difference of the two panels in Fig.\,\ref{fig: 4pcf-cartoon}. In this work, we restrict to odd $\ell_1+\ell_2+\ell_3$, and thus consider the (purely imaginary) parity-odd 4PCF.

\subsection{4PCF Estimator}
Invoking the ergodic principle, we may estimate the full 4PCF as an integral over four density fields, 
\beq\label{eq: 4pcf-naive}
    \hat\zeta(\vr_1,\vr_2,\vr_3) \equiv \frac{1}{V}\int d\vs\,\delta(\vs)\delta(\vs+\vr_1)\delta(\vs+\vr_2)\delta(\vs+\vr_3),
\eeq
where $V$ is the integration volume. This is unbiased, \textit{i.e.}\ it has expectation $\mathbb{E}[\hat\zeta] = \zeta$. Since the basis functions of \eqref{eq: basis-def} are orthonormal \citep{2020arXiv201014418C}, \eqref{eq: 4pcf-naive} can be used to construct an estimator for the 4PCF basis coefficients:
\beq\label{eq: 4pcf-basis-coeffs}
    \hat\zeta_{\ell_1\ell_2\ell_3}(r_1,r_2,r_3) &=& \int d\hr_1d\hr_2d\hr_3\,\P^*_{\ell_1\ell_2\ell_3}(\hr_1,\hr_2,\hr_3)\hat\zeta(\vr_1,\vr_2,\vr_3)\\\nonumber
    &=& \frac{1}{V}\sum_{m_1m_2m_3}\tj{\ell_1}{\ell_2}{\ell_3}{m_1}{m_2}{m_3}\\\nonumber
    &&\,\times\,\int d\vs\,d\hr_1d\hr_2d\hr_3\,\delta(\vs)\delta(\vs+\vr_1)\delta(\vs+\vr_2)\delta(\vs+\vr_3)Y_{\ell_1m_1}(\hr_1)Y_{\ell_2m_2}(\hr_2)Y_{\ell_3m_3}(\hr_3),
\eeq
using the conjugate properties of \eqref{eq: basis-conj-parity}. Defining the harmonic coefficients
\beq\label{eq: alm-def}
    \boxed{a_{\ell m}(\vs;r) \equiv \int d\hr\,\delta(\vs+\vr)Y_{\ell m}(\hr),}
\eeq
this is separable in $\hr_i$:
\beq\label{eq: 4pcf-estimator-continuous}
    \boxed{\hat\zeta_{\ell_1\ell_2\ell_3}(r_1,r_2,r_3) = \sum_{m_1m_2m_3}\tj{\ell_1}{\ell_2}{\ell_3}{m_1}{m_2}{m_3}\int \frac{d\vs}{V}\delta(\vs)a_{\ell_1m_1}(\vs;r_1)a_{\ell_2m_2}(\vs;r_2)a_{\ell_3m_3}(\vs;r_3).}
\eeq
For a discrete density field defined by $N_{\rm g}$ particles at positions $\{\vx_i\}$ with weights $w_i$, the estimator can be written as a sum:
\beq\label{eq: 4pcf-estimator-discrete}
    a_{\ell m}(\vx_i;r) &\equiv& \sum_{j=1}^{N_{\rm g}} w_jY_{\ell m}(\widehat{\vx_j-\vx_i})\delta_{\rm D}(r-|\vx_j-\vx_i|),\\\nonumber
    \hat\zeta_{\ell_1\ell_2\ell_3}(r_1,r_2,r_3) &=&  \sum_{i=1}^{N_{\rm g}}\sum_{m_1m_2m_3}\tj{\ell_1}{\ell_2}{\ell_3}{m_1}{m_2}{m_3} w_i\,a_{\ell_1m_1}(\vx_i;r_1)a_{\ell_2m_2}(\vx_i;r_2)a_{\ell_3m_3}(\vx_i;r_3),
\eeq
where the Dirac delta, $\delta_{\rm D}$, ensures that we count only secondary particles, $j$, separated from the primary, $i$, by a distance $r$. Since we must compute $a_{\ell m}$ at the location of each primary particle, the estimator requires a sum over pairs of particles, and thus has complexity $\mathcal{O}(N_{\rm g}^2)$; in practice, the scaling is closer to linear, as the $m_i$ summation is rate limiting for large $\ell_\mathrm{max}$ \citep{npcf_algo}. By replacing the Dirac function in \eqref{eq: 4pcf-estimator-discrete} by a binning function of finite width, the estimator extends to bin-averaged 4PCF estimates; we refer the reader to \citep{npcf_algo,4pcf_boss} for details. We further note that the 4PCF contains also a `disconnected' piece sourced by two copies of the 2PCF. Whilst this can be subtracted at the estimator level directly \citep{4pcf_boss}, it does not contribute to parity-odd multiplets, and will thus be ignored henceforth.

\subsection{Edge-Correction}\label{subsec: edge-correction}
Finally, estimator \eqref{eq: 4pcf-estimator-continuous} must be modified to account for the non-uniform survey geometry. For this purpose, we first define the 4PCF using the generalized Landy-Szalay form \citep{1993ApJ...412...64L,1998ApJ...494L..41S,npcf_algo}
\beq\label{eq: zeta-landy-szalay}
    \hat\zeta(\vr_1,\vr_2,\vr_3) \equiv \frac{\mathcal{N}(\vr_1,\vr_2,\vr_3)}{\mathcal{R}(\vr_1,\vr_2,\vr_3)},
\eeq
where $\mathcal{N}$ and $\mathcal{R}$ are the 4PCF estimates obtained from `data-minus-random' and random catalogs respectively, both of which are modulated by the survey window function. Following some algebra, the \textit{edge-corrected} 4PCF multiplets are given by
\beq\label{eq: edge-correction-eq}
    \zeta_{\ell_1\ell_2\ell_3}(r_1,r_2,r_3) &=& \sum_{\ell_1'\ell_2'\ell_3'}\left[M^{-1}\right]_{\ell_1\ell_2\ell_3}^{\ell_1'\ell_2'\ell_3'}(r_1,r_2,r_3)\frac{\mathcal{N}_{\ell_1'\ell_2'\ell_3'}(r_1,r_2,r_3)}{\mathcal{R}_{000}(r_1,r_2,r_3)},
\eeq
defining the coupling matrix
\beq\label{eq: edge-coupling}
    M^{\ell_1'\ell_2'\ell_3'}_{\ell_1\ell_2\ell_3}(r_1,r_2,r_3) &=& \frac{(-1)^{\ell_1'+\ell_2'+\ell_3'}}{(4\pi)^{3/2}}\sum_{L_1L_2L_3}\frac{\mathcal{R}_{L_1L_2L_3}(r_1,r_2,r_3)}{\mathcal{R}_{000}(r_1,r_2,r_3)} \left[\prod_{i=1}^3\sqrt{(2\ell_i+1)(2L_i+1)(2\ell_i'+1)}\right]\begin{Bmatrix}\ell_1 &L_1 &\ell_1' \\ \ell_2 & L_2 & \ell_2' \\ \ell_3 & L_3 & \ell_3'\end{Bmatrix}\\\nonumber
    &&\,\times \tjo{\ell_1}{L_1}{\ell_1'}\tjo{\ell_2}{L_2}{\ell_2'}\tjo{\ell_3}{L_3}{\ell_3'},
\eeq
with the curly brackets indicating a Wigner 9-$j$ symbol. This allows us to `undo' the effects of non-uniform survey geometry by measuring the 4PCF multiplets of the random field $\mathcal{R}$.\footnote{Note that this does not remove any geometry effects that couple to the \textit{anisotropic} 4PCF, nor those coupling to the 4PCF multiplets with $\ell_i>L$, assuming an initial $\ell_{\rm max}$ of $L$. The former effect is expected to be small (and usually ignored for the 3PCF \citep[e.g.,][]{2017MNRAS.469.1738S}), and the latter is ameliorated by discarding all multiplets containing $\ell_i = L$ after edge-correction, justified by noting that the coupling matrix, $M$, is close to tridiagonal.} Note that there are two manners in which an parity-odd $\zeta$ can be sourced: parity-odd $\mathcal{N}$ and parity-even $\mathcal{R}$, or parity-odd $\mathcal{R}$ and parity-even $\mathcal{N}$.\footnote{This occurs since the product of 3-$j$ symbols in the coupling matrix is zero unless $\ell_1+\ell_2+\ell_3+\ell_1'+\ell_2'+\ell_3'+L_1+L_2+L_3$ is even.} For this reason, it is imperative to restrict to parity-odd multipets only \textit{after} performing edge-correction. 

\section{Data}\label{sec: data}
\subsection{Data and Simulations}\label{subsec: data}
Our dataset comprises galaxies from the twelfth data-release (DR12) \citep{2015ApJS..219...12A} of the Baryon Oscillation Spectroscopic Survey (BOSS), part of SDSS-III \citep{2011AJ....142...72E,2013AJ....145...10D}. The survey contains two samples, CMASS and LOWZ, of which we use the former. This contains 587\,071 (216\,041) galaxies in the Northern (Southern) galactic cap (hereafter denoted NGC and SGC), across a redshift range $z\in[0.43,0.7]$ and an effective redshift of $z_\mathrm{eff} = 0.57$.\footnote{Data are publicly available at \href{https://data.sdss.org/sas/dr12/boss/lss/}{data.sdss.org/sas/dr12/boss/lss/}.} We use a fiducial cosmology $\{\Omega_m = 0.31, \Omega_bh^2 = 0.022, h = 0.676, \sigma_8 = 0.8, n_s = 0.96, \sum m_\nu = 0.06\,\text{eV}\}$ to convert angles and redshifts to Cartesian coordinates \citep[cf.\,][]{2017MNRAS.466.2242B,4pcf_boss}, and assign galaxy weights according to 
\beq\label{eq: boss-weights}
    w_\mathrm{tot} = (w_\mathrm{rf}+w_\mathrm{fc}-1)w_\mathrm{sys}w_\mathrm{fkp}.
\eeq
Here $w_\mathrm{rf}$, $w_\mathrm{fc}$, and $w_\mathrm{sys}$ correspond to redshift-failure, fiber-collision, and systematic weights respectively, with $w_\mathrm{fkp} = [1+n(z)P_0]^{-1}$ being the well-known FKP weight \citep{1994ApJ...426...23F} for background number density $n(z)$ and $P_0 = 10^4h^{-3}\mathrm{Mpc}^3$. To model the survey geometry, we use the BOSS random catalogs, containing $50\times$ more randoms than galaxies.

We additionally make use of a suite of $N_{\rm mocks} = 2048$ `MultiDark-Patchy' (hereafter \textsc{Patchy}) simulations \citep{2016MNRAS.456.4156K,2016MNRAS.460.1173R}. These are computed using an approximate gravity solver and calibrated to an $N$-body simulation, with halo occupation parameters adjusted such that the mocks well reproduce the BOSS two- and three-point statistics. These share the CMASS survey geometry and are assigned weights via
\beq
    w_\mathrm{tot} = w_\mathrm{veto}w_\mathrm{fc}w_\mathrm{fkp},
\eeq
including the veto weight $w_\mathrm{veto}$. The mocks are generated with the parameter set $\{\Omega_m = 0.3071, \Omega_bh^2 = 0.02205, h = 0.6777, \sigma_8 = 0.8288, n_s = 0.96, \sum m_\nu = 0\,\text{eV}\}$ and coordinates are converted using the BOSS fiducial cosmology.

\subsection{4PCF Estimates}\label{subsec: data-4pcf}
One of the main drawbacks with higher-order NPCFs is their dimensionality. To characterize the 4PCF, we must specify three multiplet indices ($\ell_1,\ell_2,\ell_3$) and three radial bins ($r_1,r_2,r_3$), which can lead to a statistic with a large number of (highly correlated) elements \citep{npcf_algo}. For this reason, we adopt a relatively coarse radial binning scheme using $N_r = 10$ linearly spaced radial bins in $[20, 160]\Mpch$, giving $\Delta r = 14\Mpch$. Furthermore, we enforce $r_2>r_1+\Delta r$ and $r_3>r_2+\Delta r$, to ensure that the the separation between any two galaxies in the 4PCF tetrahedron is at least $\Delta r$ (cf.\,Fig.\,\ref{fig: 4pcf-cartoon}). This removes modes from the non-linear region; these are difficult to model and can be strongly affected by baryonic physics. For the angular binning, we fix $\ell_\mathrm{max} = 5$, leading to a total of $56$ radial bins and $111$ multiplets (both parity-odd and parity-even), hence $6\,216$ elements in the full 4PCF statistic. In the analysis of \S\ref{sec: analysis-methods}, we use only the $23$ multiplets with odd $\ell_1+\ell_2+\ell_3$ and $\ell_i\leq 4$, giving a total of $N_\zeta = 1288$ elements; the rest are required for edge-correction (\S\ref{subsec: edge-correction}).

Computation of the 4PCF multiplets, $\zeta_{\ell_1\ell_2\ell_3}(r_1,r_2,r_3)$, is performed using the \textsc{encore} code \citep{npcf_algo}. We separately measure the contributions from a random catalog and a set of 32 `data-minus-random' catalogs, each with $1.5\times$ the galaxy density; the latter are averaged to form the $\mathcal{N}$ quantities entering the edge-correction equation \eqref{eq: zeta-landy-szalay}, whilst the former give $\mathcal{R}$.\footnote{If the algorithm's runtime scales as $N_{\rm g}^2$, this partitioning minimizes the Poisson error at fixed computational cost \citep{2015MNRAS.454.4142S,2020JCAP...12..021S}. In our case, the scaling is closer to linear, thus the total work is roughly independent of the partition size.} Using \eqref{eq: edge-correction-eq}, the quantities are then  combined to form the edge-corrected 4PCF multipoles.

\begin{figure}
    \centering
    \includegraphics[width=0.9\textwidth]{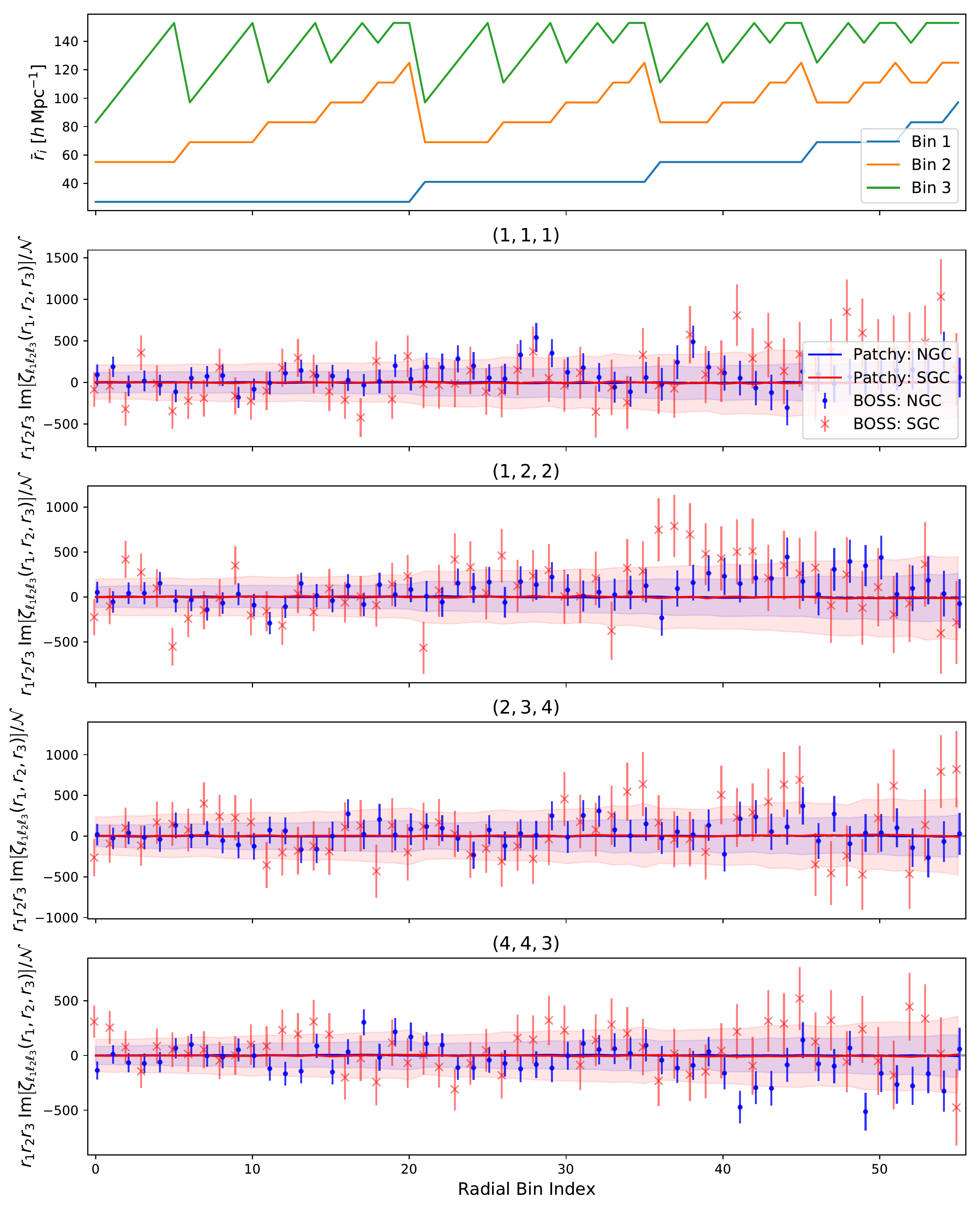}
    \caption{Measurements of the parity-odd 4PCF from the BOSS CMASS galaxy sample, alongside those from 2048 \textsc{Patchy} simulations. The NGC (SGC) results are shown in blue (red) bands, with the BOSS data shown as error-bars, using the \textsc{Patchy} variances. Results are displayed for a selection of $\{\ell_1,\ell_2,\ell_3\}$ multiplets (which specify the internal angles of the galaxy tetrahedron, as in Fig.\,\ref{fig: 4pcf-cartoon}), whose values are indicated by the title of each subfigure. In total, 23 parity-odd multiplets are included in the analysis of \S\ref{sec: results}. The horizontal axis specifies the radial bin combinations, $\{r_1,r_2,r_3\}$, with the central values of $r_1$, $r_2$ and $r_3$ in each bin shown in the top panel. These correspond to the distances of the secondary, tertiary, and quaternary galaxies from the primary in Fig.\,\ref{fig: 4pcf-cartoon}. For visibility, the 4PCF measurements are rescaled by a factor $-i\,r_1r_2r_3$. As expected, the \textsc{Patchy} measurements show no signs of parity-violation. \unblind{Given the high correlation between neighboring bins, it is difficult to visually assess whether the BOSS dataset contains signatures of parity-violation;} this is quantified in Figs.\,\ref{fig: rank-test}\,\&\,\ref{fig: projected-results}. }
    \label{fig: 4pcf-results}
\end{figure}

For samples with similar number densities to BOSS, the runtime of \textsc{encore} scales as $N_{\rm g}N_r^3(\ell_\mathrm{max}+1)^5$ \citep{npcf_algo}, with computation dominated by the $m_i$ summations of \eqref{eq: 4pcf-estimator-discrete} rather than estimation of the harmonic coefficients $a_{\ell m}$ (which scales as $N_{\rm g}^2(1+\ell_\mathrm{max})^2$, albeit with a more modest prefactor). In practice, we parallelize computation using \textsc{OpenMP}, with each NGC (SGC) each simulation requiring $\sim$\,$32$\,($6$) CPU-hours to analyze on a modern 16-core Intel processor, including edge-correction. In total, analysis of the BOSS data and 2048 \textsc{Patchy} mocks required $\sim$\,$80$k\,CPU-hours. This is comparable to the computational costs of the 2PCF analysis in Ref.\,\citep{2018MNRAS.477.1153V}, and is facilitated by the efficient nature of the \textsc{encore} algorithm. We display a selection of the measured 4PCF multiplets in Fig.\,\ref{fig: 4pcf-results}.

\subsection{Covariance Matrices}\label{subsec: analytic-cov}

The \textsc{Patchy} mocks described in \S\ref{subsec: data} can be used to form a sample covariance of the 4PCF statistic in the standard manner:
\beq\label{eq: sample-4pcf-cov}
    \hat{\mathsf{C}}_{\ell_1^{}\ell_2^{}\ell_3^{};\ell_1'\ell_2'\ell_3'}(r_1^{},r_2^{},r_3{};r_1',r_2',r_3')  &=& \frac{1}{N_{\rm mocks}-1}\sum_{i=1}^{N_{\rm mocks}} \left(\zeta^{(i)}_{\ell_1^{}\ell_2^{}\ell_3^{}}(r_1^{},r_2^{},r_3^{})-\bar{\zeta}_{\ell_1^{}\ell_2^{}\ell_3^{}}(r_1^{},r_2^{},r_3^{})\right)\\\nonumber
    &&\qquad\qquad\qquad\,\times\,\left(\zeta^{(i)}_{\ell_1'\ell_2'\ell_3'}(r_1',r_2',r_3')-\bar{\zeta}_{\ell_1'\ell_2'\ell_3'}(r_1',r_2',r_3')\right),
\eeq
where $\zeta^{(i)}$ represents the $i$-th 4PCF estimate (in the NGC or SGC region), and $\bar{\zeta}$ is the average over $N_{\rm mocks}$ realizations. Since the number of 4PCF bins exceeds the number of \textsc{Patchy} mocks, this is not invertible, making it difficult to perform traditional $\chi^2$-based analyses. For this reason, we supplement the sample covariance with the analytic covariance described in \citep{npcf_cov}. Essentially, this computes:
\beq\label{eq: covariance}
    \mathrm{Cov}(\vr_1^{},\vr_2^{},\vr_3^{};\vr_1',\vr_2',\vr_3') &=& \int \frac{d\vs}{V}\frac{d\vs'}{V}\av{\delta(\vs)\delta(\vs+\vr_1)\delta(\vs+\vr_2)\delta(\vs+\vr_3)\delta(\vs')\delta(\vs'+\vr_1')\delta(\vs'+\vr_2')\delta(\vs'+\vr_3')}\\\nonumber
    &&\,-\,\int \frac{d\vs}{V}\av{\delta(\vs)\delta(\vs+\vr_1)\delta(\vs+\vr_2)\delta(\vs+\vr_3)}\int \frac{d\vs'}{V}\av{\delta(\vs')\delta(\vs'+\vr_1')\delta(\vs'+\vr_2')\delta(\vs'+\vr_3')},
\eeq
where the statistical expectations can be expanded using Wick's theorem to yield products of four 2PCFs. The covariance is then projected into the angular basis of \S\ref{subsec: basis} and simplified. The approach makes a number of assumptions:
\begin{itemize}
    \item \textbf{Isotropy}: The 2PCF $\xi(\vr)\equiv\av{\delta(\vs)\delta(\vs+\vr)}$ is assumed to be a function only of $|\vr|$. This neglects redshift-space distortions, which have a non-trivial impact on the isotropic 4PCF covariance.
    \item \textbf{Gaussianity}: The expectations entering \eqref{eq: covariance} strictly contain additional contributions from higher-order correlators such as the 3PCF.
    \item \textbf{Survey Geometry}: Whilst the 4PCF is edge-corrected (\S\ref{subsec: edge-correction}), the same is not true for the covariance. The latter inherits non-trivial dependence on the survey geometry \citep[e.g.,][]{2020PhRvD.102l3517W,2019MNRAS.490.5931P}, which cannot be simply captured by modifying the survey volume or shot-noise \citep{npcf_cov}.
\end{itemize}
For these reasons, we do \textit{not} expect the analytic models of \citep{npcf_cov} to accurately predict the true covariance of BOSS. It is a relatively close approximation of matrix structure however, and will thus be used as a proxy covariance to facilitate the analysis techniques described in \S\ref{sec: analysis-methods}. We construct the covariance using the same radial binning parameters as in \S\ref{subsec: data}, restricting to odd $\ell_1+\ell_2+\ell_3$. Following the prescription of \citep{2020PhRvD.102l3517W} (but generalized to higher dimensions), we use an effective volume of $1.90h^{-3}\mathrm{Gpc}^3$ ($0.77h^{-3}\mathrm{Gpc}^3$) and shot-noise $P_\mathrm{shot} = 3130h^{-3}\mathrm{Mpc}^3$ ($3160h^{-3}\mathrm{Mpc}^3$) for the NGC (SGC) subsample. The input 2PCFs are taken from a fit to the BOSS CMASS power spectrum, modelled using the Effective Field Theory of Large Scale Structure \citep{2020JCAP...05..042I}, as implemented in \textsc{class-pt} \citep{2020PhRvD.102f3533C}.

\begin{figure}[!ht]
    \centering
    \subfloat[Correlation Matrices\label{fig: cov-corr}]{%
    \includegraphics[width=0.9\textwidth]{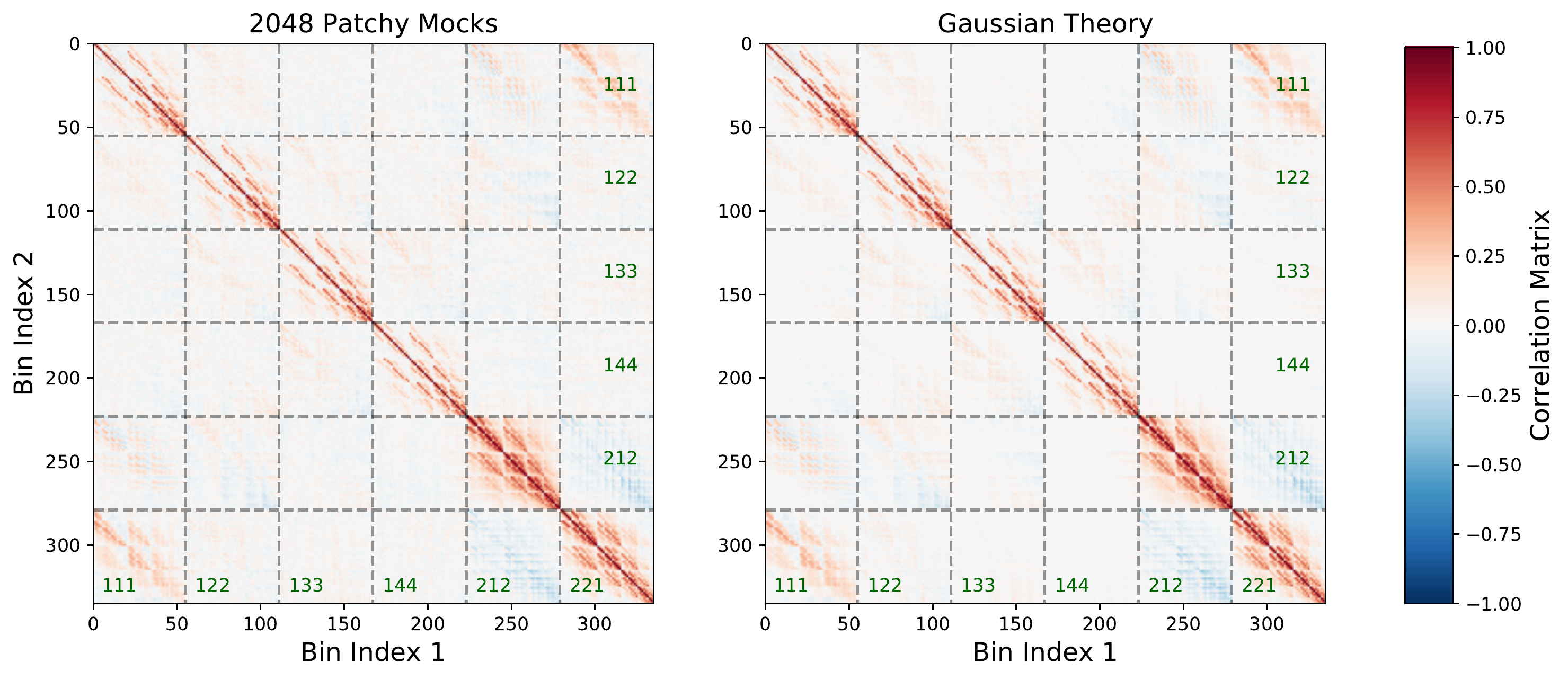}
    }
    \hfill
    \subfloat[Covariance Diagonal\label{fig: cov-diag}]{%
        \includegraphics[width=0.6\textwidth]{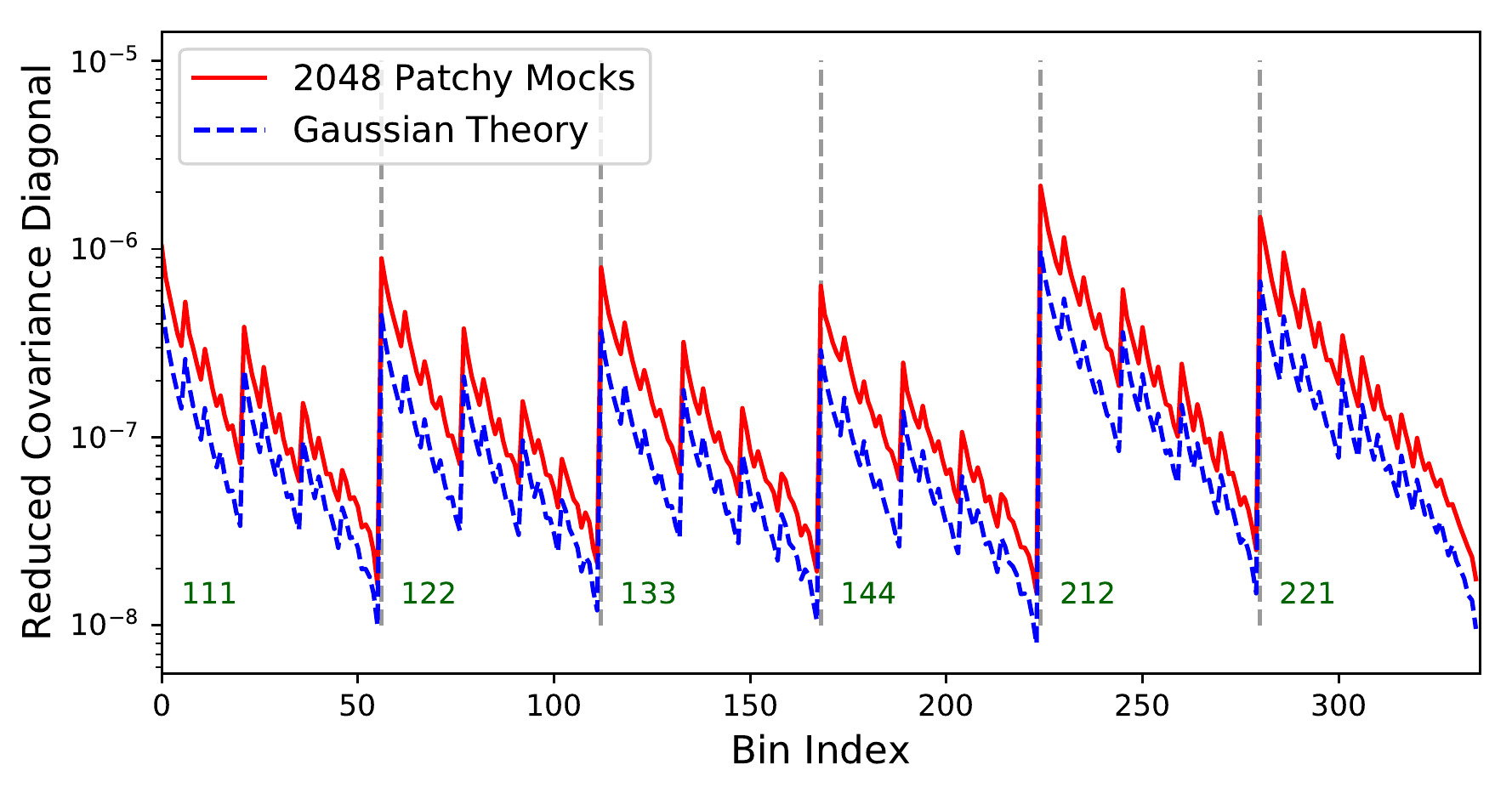}
    }
    \caption{Comparison of the sample and analytic covariance matrices for the parity-odd 4PCF of the BOSS CMASS NGC region. The former are estimated using \eqref{eq: sample-4pcf-cov}, whilst the latter use the approach of \citep{npcf_cov}, which does not include redshift-space distortions, non-Gaussianity, or the effects of survey geometry. Fig.\,\ref{fig: cov-corr} compares the correlation matrices (defined as the covariance matrices normalized by their diagonals); we see similar structure in both cases. The rows and columns represent the indices of the 4PCF, collapsed into one dimension, with each submatrix (indicated by the dotted lines) showing a different multiplet $\{\ell_1,\ell_2,\ell_3\}$, as labelled in green. Elements within a submatrix are ordered in increasing radii $r_1,r_2,r_3$. We include only the first six multiplets here; 23 are used in the analysis of \S\ref{sec: results}\,\&\,\ref{subsec: CS-results}. Fig.\,\ref{fig: cov-diag} shows the corresponding diagonal elements of the covariance. Notably, the analytic covariance is an underestimate by a factor close to two; we expect this to arise primarily due to the non-uniform survey geometry of the CMASS region \citep{npcf_cov}.}\label{fig: 4pcf-cov}
\end{figure}

Fig.\,\ref{fig: 4pcf-cov} compares the analytic and sample covariances for the NGC region, with the latter estimated from \eqref{eq: sample-4pcf-cov} using 2048 \textsc{Patchy} mocks. Considering the correlation matrices (Fig.\,\ref{fig: cov-corr}, defined as $\mathsf{R}_{ij} \equiv \mathsf{C}_{ij}/\sqrt{\mathsf{C}_{ii}\mathsf{C}_{jj}}$ for covariance $\mathsf{C}_{ij}$), we find good agreement between the two, indicating that the Gaussian theory model well reproduces the matrix structure. However, the diagonal elements (Fig.\,\ref{fig: cov-diag}) of the analytic covariance are roughly a factor of two less than those of the sample covariance. This is likely to arise from the non-trivial survey geometry of the BOSS CMASS region \citep{npcf_cov} and prohibits direct use of the analytic covariance as a model for the 4PCF statistics.\footnote{Note that this discrepancy is \textit{not} fully resolved by rescaling the theory covariance by a constant factor.}

\section{Analysis Methods}\label{sec: analysis-methods}
Below, we discuss two techniques that will be used to search for a signature of parity-violation in \S\ref{sec: results}: (1) a non-parametric rank test, which does not require the likelihood to be Gaussian, and (2) data compression followed by a mock-based $\chi^2$-analysis. Both approaches make use of the smooth (but inaccurate) covariance matrix model of \S\ref{subsec: analytic-cov} to overcome the difficulties associated with the high-dimensionality of the 4PCF. To avoid confirmation bias, the pipeline implementing these techniques\footnote{Available at \href{https://github.com/oliverphilcox/Parity-Odd-4PCF}{github.com/oliverphilcox/Parity-Odd-4PCF}.} was constructed before the BOSS data were unblinded.

\subsection{Non-Parametric Rank Test}\label{subsec: rank-test}
Non-parametric tests provide a powerful way to analyze data when the underlying likelihood is not known. Here, we consider a \textit{rank test}, examining the null hypothesis of zero parity-odd 4PCF. To implement this, we first define a test statistic, computed on both the data and a set of mocks. These simulations are required to obey the null hypothesis (\textit{i.e.}\ be parity-invariant) and have realistic noise properties. The test statistic measured from data is then compared to the empirical distribution obtained from the mocks, allowing construction of a detection significance. For example, if the data statistic exceeds that of $95\%$ of the mocks, we may reject the null hypothesis at $95\%$ CL. The principal advantage of this approach is that it does not require a theoretical PDF for the test statistic, \textit{i.e.}\ we do not have to assume the 4PCF to be a draw from some multivariate Gaussian. Indeed, the observed 4PCF does \textit{not} appear to be Gaussian; this is explored in Appendix \ref{appen: NG-likelihood}. A limitation of such rank tests is that one cannot claim a detection at high significance; rather the maximal confidence level is $(1-1/N_{\rm mocks})$. 

Below, we will use the following test statistic, dubbed the \textit{pseudo}-$\chi^2$:
\beq\label{eq: pseudo-chi2}
    \tilde{\chi}^2 \equiv \left[\zeta^T \tilde{\mathsf C}^{-1} \zeta\right]_{\rm NGC} + \left[\zeta^T \tilde{\mathsf C}^{-1} \zeta\right]_{\rm SGC},
\eeq
where $\zeta$ is the set of measured parity-odd 4PCF multipoles (treated as a $N_\zeta$-dimensional vector), and $\tilde{\mathsf{C}}$ is the theoretical covariance matrix (\S\ref{subsec: analytic-cov}). If $\tilde{\mathsf{C}}$ is equal to the sample covariance (in the limit of infinite mocks), \eqref{eq: pseudo-chi2} reduces to the usual $\chi^2$ statistic, given a fiducial model of zero parity-odd 4PCF and assuming the NGC and SGC regions to be independent. Whilst the covariances are not quite equal in practice (Fig.\,\ref{fig: 4pcf-cov}), we expect \eqref{eq: pseudo-chi2} to produce a close-to-optimal weighting for the data, particularly if the likelihood is close to Gaussian. Furthermore, since the pseudo-$\chi^2$ statistic does not subtract off a mean, the rank test will naturally account for any spurious parity-odd contributions that are present in both \textsc{Patchy} and BOSS. These might arise from imperfections in the edge-correction routine or lightcone projection effects. To perform the test, we simply compute $\tilde{\chi}^2$ for BOSS and each of the $N_{\rm mocks}=2048$ \textsc{Patchy} simulations (\S\ref{subsec: data}), before assigning a detection significance from the empirical \textsc{Patchy} PDF. 


\subsection{Compressed Gaussian Analysis}\label{subsec: compressed-analysis}
A common trick when dealing with high-dimensional statistics is to apply some form of data compression \citep[e.g.,][]{2000MNRAS.317..965H,2000ApJ...544..597S,2018MNRAS.476L..60A,2021PhRvD.103d3508P}. In general, this proceeds by projecting the data onto some (small) set of basis vectors, thus greatly reducing the dimensionality. When performing parameter inference, basis vectors are usually chosen to preserve the Fisher information matrix \citep[e.g.,][]{2000MNRAS.317..965H,2018MNRAS.476L..60A} or the log-likelihood \citep{2021PhRvD.103d3508P}. Since our primary goal in this work is to search for signatures of parity-violation in a model-agnostic fashion, we adopt a somewhat different compression scheme, following \citep{2000ApJ...544..597S,4pcf_boss}.

Here, we project the 4PCF onto a basis given by the eigenvectors of the theoretical covariance matrix (\S\ref{subsec: analytic-cov}). Explicitly, we define the projected statistic
\beq\label{eq: compressed-statistic}
    v \equiv \mathsf{U}^T\zeta,
\eeq
where the orthogonal matrix $\mathsf{U}$ is specified by $\tilde{\mathsf{C}} = \mathsf{U}\mathsf{\Lambda}\mathsf{U}^T$ for diagonal eigenvalue matrix $\mathsf{\Lambda}$. The compressed statistic has covariance $\expect{vv^T} = \mathsf{U}^T\mathsf{C}\mathsf{U}$, where $\mathsf{C}$ is the covariance of $\zeta$; if the theory and analytic covariances agree, this is diagonal and equal to $\L$. In practice, we expect the compressed coefficients to be almost independent.

To perform dimensionality reduction, we must restrict to a subset of the aforementioned basis vectors. Given that we have no prior on the shape of a parity-violating 4PCF signal, we cannot select the basis vectors based on signal-to-noise considerations (as in \citep{2000ApJ...544..597S,4pcf_boss}). 
Instead, we use the $N_{\rm eig}$ eigenvectors with smallest $\Lambda_i$, corresponding to the directions that can be most well measured.\footnote{Additional choices of basis functions can be found in \S\ref{subsec: sys-compression}.} This highlights the benefits of using the \textit{theoretical} covariance matrix to perform the projection; since the sample covariance does not have full rank, its smallest eigenvalues are not well defined.

Following selection of the basis vectors, we project both the BOSS data and the \textsc{Patchy} mocks into the $N_{\rm eig}$-dimensional subspace using \eqref{eq: compressed-statistic}. As in \eqref{eq: sample-4pcf-cov}, we can form a sample covariance for $v$ from the \textsc{Patchy} measurements:
\beq\label{eq: compressed-sample-cov}
    \hat{\mathsf{C}}_{v,\alpha\beta} = \frac{1}{N_{\rm mocks}-1}\sum_{i=1}^{N_{\rm mocks}} \left(v^{(i)}_\alpha-\bar{v}^{}_\alpha\right)\left(v^{(i)}_\beta-\bar{v}^{}_\beta\right), 
\eeq
where $v_\alpha^{(i)}$ indicates the compressed 4PCF of the $i$-th mock, with $\alpha,\beta\in\{1,\ldots,N_{\rm eig}\}$. Assuming $N_{\rm mocks}>N_{\rm eig}$, the sample covariance has full rank (unlike the uncompressed 4PCF covariance), and is thus invertible. In the low-dimensional subspace, analysis centers around the following statistic:
\beq\label{eq: T2-def}
    \hat{T}^2 = v^T\hat{\mathsf{C}}_v^{-1}v,
\eeq
where we have assumed zero mean, as in the null hypothesis. If $v$ is assumed to be Gaussian distributed (a fair assumption if the dimensionality is small), $\hat{T}^2$ follows a $\chi^2$-distribution with $N_{\rm eig}$ degrees of freedom in the limit of large $N_{\rm mocks}$. In practice, we must account for noise in the sample covariance $\hat{\mathsf{C}}_v$. A approach is to add the `Hartlap' correction factor \citep{1933PCPS...29..260W,2007A&A...464..399H}, leading to the modified statistic
\beq\label{eq: H2-def}
    \hat{H}^2 = f_H\times v^T\hat{\mathsf{C}}_v^{-1}v, \qquad f_H = \frac{N_{\rm mocks}-N_{\rm eig}-2}{N_{\rm mocks} - 1},
\eeq
whose expectation is $\chi^2$.
$\hat{H}^2$ is then analyzed using a $\chi^2$-distribution, assuming Gaussianity. However, this does not correctly treat the sample covariance noise, and results in a PDF which is too sharply peaked if $N_{\rm eig}$ is close to $N_{\rm mocks}$ \citep{2016MNRAS.456L.132S}. Instead, one should analyze the $\hat{T}^2$ statistic \eqref{eq: T2-def} directly, using the PDF:
\beq\label{eq: T2-pdf}
    f_T(T^2;n,p) = \frac{\Gamma\left[(n+1)/2\right]}{\Gamma(p/2)\Gamma\left[(n-p+1)/2\right]}\frac{n^{-p/2}(T^2)^{p/2-1}}{(T^2/n+1)^{(n+1)/2}},
\eeq
where $n = N_{\rm mocks}-1$, $p=N_{\rm eig}$, and $\Gamma$ is the Gamma function \citep{2016MNRAS.456L.132S}. When dealing with multiple independent datasets (\textit{i.e.}\ the NGC and SGC 4PCF measurements), one may sum the two $T^2$ estimates; the resulting PDF is the convolution of two copies of \eqref{eq: T2-pdf}, and is easily evaluated with a Fast Fourier Transform. 
This approach will be adopted for the main analysis of \S\ref{sec: results} to ensure that we do not claim a false detection of parity-violation.


Finally, we comment on the validity of our compression scheme. By the Eckart-Young theorem \citep{eckart-young}, the scheme is optimal (in terms of inverse-variance) in the limit of $\tilde{\mathsf{C}} = \mathsf{C}$ and a Gaussian covariance. Since the data and mocks are compressed in the same manner, it is unbiased for any choice of projection matrix $\mathsf{U}$ or dimension $N_{\rm eig}$. If too few basis vectors are used or if the theory covariance is far from the truth, the penalty (in the limit of large $N_{\rm mocks}$) is simply a reduced detection significance.\footnote{This is easiest to show by considering the average $\chi^2$ difference between some signal $\zeta_0$ and the null hypothesis of $\expect{\hat\zeta} = 0$. Without compression, $\Delta\chi^2 = \zeta_0^T\mathsf{C}^{-1}\zeta_0$, whilst following projection by some $N_\zeta\times N_{\rm eig}$ matrix $\mathsf{U}$, $\Delta\chi^2_{\rm proj} = \zeta_0^T\mathsf{U}\left(\mathsf{U}^T\mathsf{C}\mathsf{U}\right)^{-1}\mathsf{U}^T\zeta_0$. If the projection is optimal, \textit{i.e.}\ if $\mathsf{U}$ is the eigenvector matrix of $\mathsf{C}$, then $\Delta\chi^2_{\rm proj, opt} = \sum_{i=1}^{N_{\rm eig}} \bar\zeta_{0,i}^2/\Lambda_i$ where $\bar\zeta_0 = \mathsf{U}^T\zeta_0$. Since $\bar\zeta_{0,i}^2\geq 0$ and $\Lambda_i>0$, it is clear that $\Delta\chi^2_{\rm proj, opt}\leq \Delta\chi^2$, with equality iff $N_{\rm eig} = N_\zeta$. A similar result holds in the more general case due to the properties of projection matrices; this is easily shown by rotating to a diagonal basis.} When using finite $N_{\rm mocks}$, increasing $N_{\rm eig}$ will also lead to increased noise in the covariance matrix $\hat{\mathsf{C}}_v$, somewhat lessening the detection significance. To incorporate these effects, a range of $N_{\rm eig}$ values will be considered in \S\ref{sec: results}. 

\section{Null-Test Results}\label{sec: results}
We now proceed to assess whether the BOSS data contains signatures of parity-violation. Firstly, we consider the raw 4PCF measurements, displayed in Fig.\,\ref{fig: 4pcf-results} for a selection of multiplets $\{\ell_1,\ell_2,\ell_3\}$. As expected, the mean parity-odd 4PCF of \textsc{Patchy} mocks appears close to zero. This functions as a useful consistency test for the analysis pipeline; errors in the 4PCF computation could have led to a detectable signal \unblind{in \textsc{Patchy} mocks}, especially given that the parity-even 4PCF is large \citep{4pcf_boss}. For the standard-deviations, we find a rough scaling of $(r_1r_2r_3)^{-1}$, with enhanced amplitudes found for the SGC region due to its smaller volume (\S\ref{sec: data}). Moving to the BOSS results, we find considerable (highly correlated) scatter around zero, but no obvious signatures of parity-violation.

\begin{figure}
    \centering
    \includegraphics[width=0.45\textwidth]{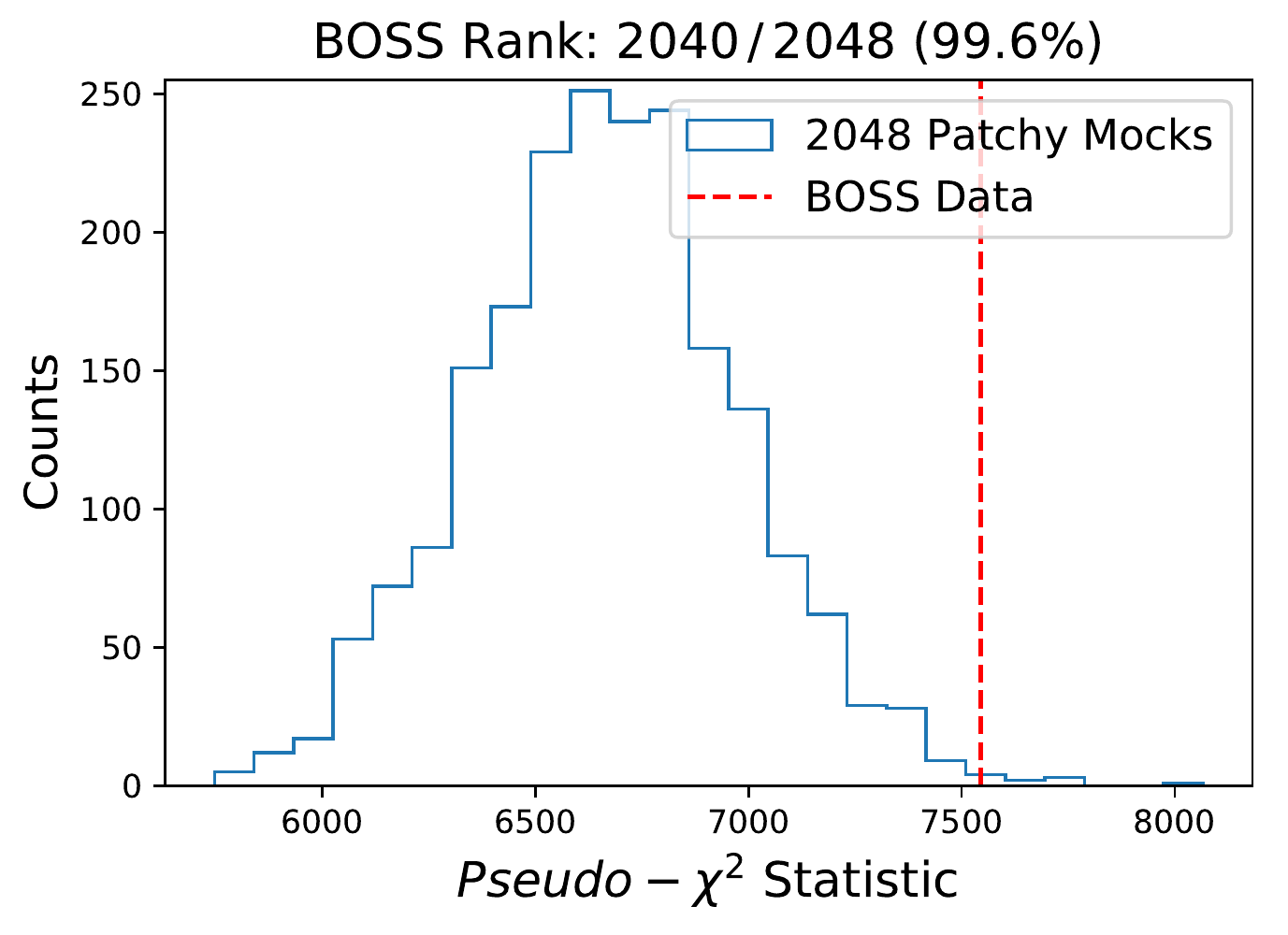}
    \caption{Empirical distribution of the \textit{pseudo}-$\chi^2$ test statistic defined in \eqref{eq: pseudo-chi2} from 2048 \textsc{Patchy} mocks and the BOSS data. This is a proxy for the standard $\chi^2$ parameter, but uses the theoretical covariance matrix of \S\ref{subsec: analytic-cov}. The data (shown as a vertical dashed line) has a CDF of \unblind{$99.6\%$; this is inconsistent
    with the null hypothesis of parity-invariance at $2.9\sigma$.} Note that this test does not assume Gaussianity of the likelihood, and naturally encompasses any spurious parity-odd contributions appearing in both the \textsc{Patchy} mocks and the BOSS data. 
    \unblind{The significant detection of parity-violation found by this test indicates either parity-violating physics or unresolved systematics.} The above plot represents the main result of this work.}
    \label{fig: rank-test}
\end{figure}

To examine this further, we turn to the non-parametric rank test of \S\ref{subsec: rank-test}. In Fig.\,\ref{fig: rank-test}, we plot the empirical distribution of the \textit{pseudo}-$\chi^2$ test statistic; this appears to have broader tails than a standard $\chi^2$ distribution (most likely due to imperfections in the theoretical covariance $\tilde{\mathsf{C}}$), highlighting the importance of a non-parametric treatment. The BOSS data has a rank of \unblind{$2040/2048$}, 
and an associated \unblind{\textbf{detection probability of $99.6\%$ (equivalent to $2.9\sigma$)}. This is inconsistent with a random draw from the empirical distribution of \textsc{Patchy} mocks, and gives \textbf{evidence for parity-violation}. A greater number of simulations would be needed to probe this detection at higher significance.}

\begin{figure}
    \centering
    \includegraphics[width=\textwidth]{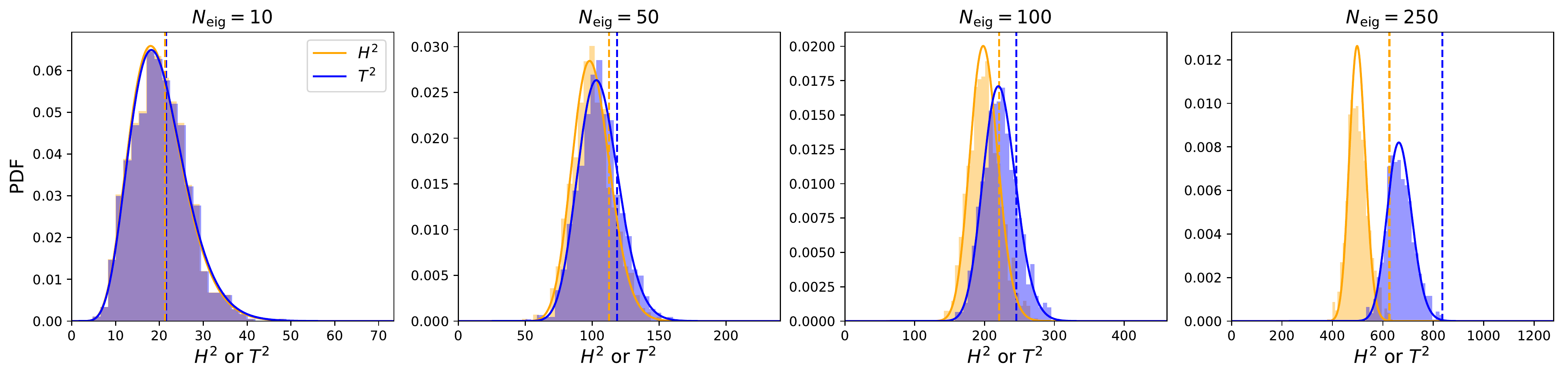}\\
    \includegraphics[width=\textwidth]{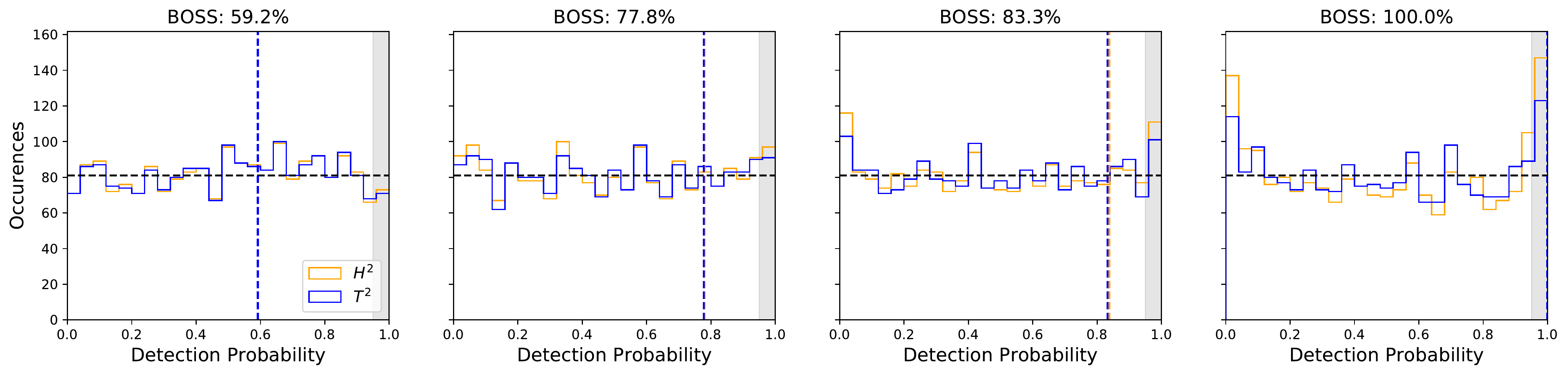}
    \caption{Distributions of the $H^2$ and $T^2$ statistics \eqref{eq: H2-def}\,\&\,\eqref{eq: T2-def} for the compressed BOSS data and \textsc{Patchy} simulations. We show results using various numbers of basis vectors, $N_{\rm eig}$, as indicated by the titles, and note that the statistic includes both NGC and SGC measurements. The top panels compare the theoretical and empirical PDFs for each statistic, whilst the bottom panels display the CDFs. Results for BOSS are shown as vertical dashed lines in both cases. To compute the empirical distributions (shown as histograms), we apply bootstrapping; $(N_{\rm mocks}-1)$ mocks are used to define a sample covariance, allowing computation of $H^2$ and $T^2$ for the excluded mock. The theoretical PDFs for $H^2$ and $T^2$ are the convolution of two $\chi^2$ or $T^2$ distributions \eqref{eq: T2-pdf}. 
    For BOSS, we report detection probabilities of 
    \unblind{$59.2\%$, $77.8\%$, $83.3\%$, and $100.0\%$}
    from the $T^2$ statistic using 
    \unblind{$N_{\rm eig} = 10$, $50$, $100$, and $250$} basis vectors respectively. \unblind{As in Fig.\,\ref{fig: rank-test}, we find mild evidence for parity-violation, particularly as $N_{\rm eig}$ increases. We caution that this test assumes a Gaussian likelihood, which may lead to overestimated detection probabilities at high $N_{\rm eig}$ (as suggested by the somewhat skewed empirical distribution of \textsc{Patchy} mocks at $N_{\rm eig}=250$, and the results of Appendix \ref{appen: NG-likelihood}, which correspond to $N_{\rm eig} = N_{\zeta}$).}}
    \label{fig: projected-results}
\end{figure}

An additional test is given by the projected $\chi^2$-based analysis discussed in \S\ref{subsec: compressed-analysis}. Whilst this assumes a Gaussian distribution for the compressed statistic (somewhat justified by the reduced number of bins), it uses the sample covariance rather than the theoretical covariance, and thus optimally weights the data. In the top panel of Fig.\,\ref{fig: projected-results}, we show the theoretical and empirical distributions of the projected sample statistics $T^2$ and $H^2$ (\S\ref{subsec: compressed-analysis}) from \textsc{Patchy}, with the latter obtained via bootstrapping. \unblind{For small $N_{\rm eig}$, the empirical distributions of both statistics}
seem well-fit by their theory models, which is expected since $N_{\rm mocks}$ is considerably larger than $N_{\rm eig}$.
\unblind{At larger $N_{\rm eig}$ we begin to see discrepancies between the $H^2$ statistic and the accompanying $\chi^2$ theory model, with the former having a slightly narrower distribution. Considering the detection CDFs shown in the lower panel of Fig.\,\ref{fig: projected-results}, the distribution of \textsc{Patchy} mocks appear somewhat non-uniform for the $H^2$ statistic at $N_{\rm eig}>100$.} In particular, \unblind{152} (\unblind{199}) mocks lie in the outermost $5\%$ of the theory distribution for \unblind{$N_{\rm eig} = 100$} (\unblind{$N_{\rm eig} = 250$}), in contrast to the $102\pm10$ expected. This echoes the conclusion of \citep{2016MNRAS.456L.132S}; \unblind{if $N_{\rm eig}$ is close to $N_{\rm mocks}$, improper treatment of covariance matrix noise may be dangerous, and could lead to false detections of parity-violation. For the $T^2$ distribution (which correctly treats such effects) we find somewhat better agreement, with $135$ and $162$ mocks in the outer 5\% region respectively. However, the distribution still appears to be somewhat skewed. We attribute this to intrinsic non-Gaussianity of the 4PCF likelihood (see \citep[e.g.,][]{2019MNRAS.485.2956H} and Appendix \ref{appen: NG-likelihood}), whose effect increases as the size of the data vector increases, and the Central Limit Theorem becomes less applicable.} 

\unblind{For the BOSS data, the compressed Gaussian analysis gives detection significances of} \unblind{$59.2\%$, $77.8\%$, $83.3\%$, and $100.0\%$
for $N_{\rm eig} = 10, 50, 100$, and $250$ respectively, equivalent to $1.3\sigma, 1.7\sigma$, $1.9\sigma$, and $3.9\sigma$ in a two-tail test. The fact that these results are a strong function of $N_{\rm eig}$ suggests that our projection scheme is inefficient, \textit{i.e.}\ that 4PCF components dropped from the analysis carry significant information regarding the parity-violating signature.\footnote{This differs from the conclusion of \citep{4pcf_boss}, which used a similar compression scheme to analyze the parity-even 4PCF. In the former work, basis vectors were chosen based on the mean non-Gaussian signal from the mocks, ensuring optimal linear compression. This is not possible in our case, since the mocks conserve parity.} This is a consequence of performing a blind analysis; given the lack of a physical model, we cannot choose basis vectors which maximize the signal-to-noise (though see \S\ref{subsec: sys-compression} for results using alternative choices of compression scheme). Overall, we find a weak preference for a non-Gaussian signal using the compressed analysis. This notwithstanding, we again find a preference for a non-zero parity-odd signal from the sample, though again caution that the $N_{\rm eig}=250$ result may be artificially enhanced by likelihood non-Gaussianity.}

\unblind{To close, we comment on the implications of the results found herein. Firstly, we note that the results are largely consistent between the two tests, with both finding a detection of large-scale parity-violation at $\sim$\,$3\sigma$. This cannot be caused by an incorrectly modelled likelihood (evidenced by the rank test), nor is it a result of our analysis incorrectly treating the window function (which would have led to the compressed analysis showing a parity-violating signature in the \textsc{Patchy} mocks). \textbf{This leaves two explanations: (1) new physics, and (2) unexplained systematics.} Whilst the first is a distinct possibility (and is not ruled out by other observations, since no former experiment has measured the 4PCF in a model-agnostic fashion), the second should also be carefully considered. A brief discussion of this is presented in the next section.}

\section{Systematic Tests}\label{sec: systematics}

Below, we report the results of various checks performed to test the results of \S\ref{sec: results}, utilizing data cuts, mock catalogs, and rescaled statistics. In the final section, we will also comment on potential sources of systematic effects that could lead to a false detection of parity-violation. All the analyses below were devised post-unblinding. 

\subsection{Data Partitioning}\label{subsubsec: sys-ngc-sgc}
A simple test for systematic errors is to repeat the non-parametric rank test described in \S\ref{subsec: rank-test} for the two BOSS regions (NGC and SGC) separately. The two patches are in opposite hemispheres, have different calibrations, and are of different angular sizes, thus this is a sensitive test of the effects of survey calibration and large-scale modes.  

\begin{figure}
\centering
\begin{minipage}{.55\textwidth}
  \centering
  \includegraphics[width=\linewidth]{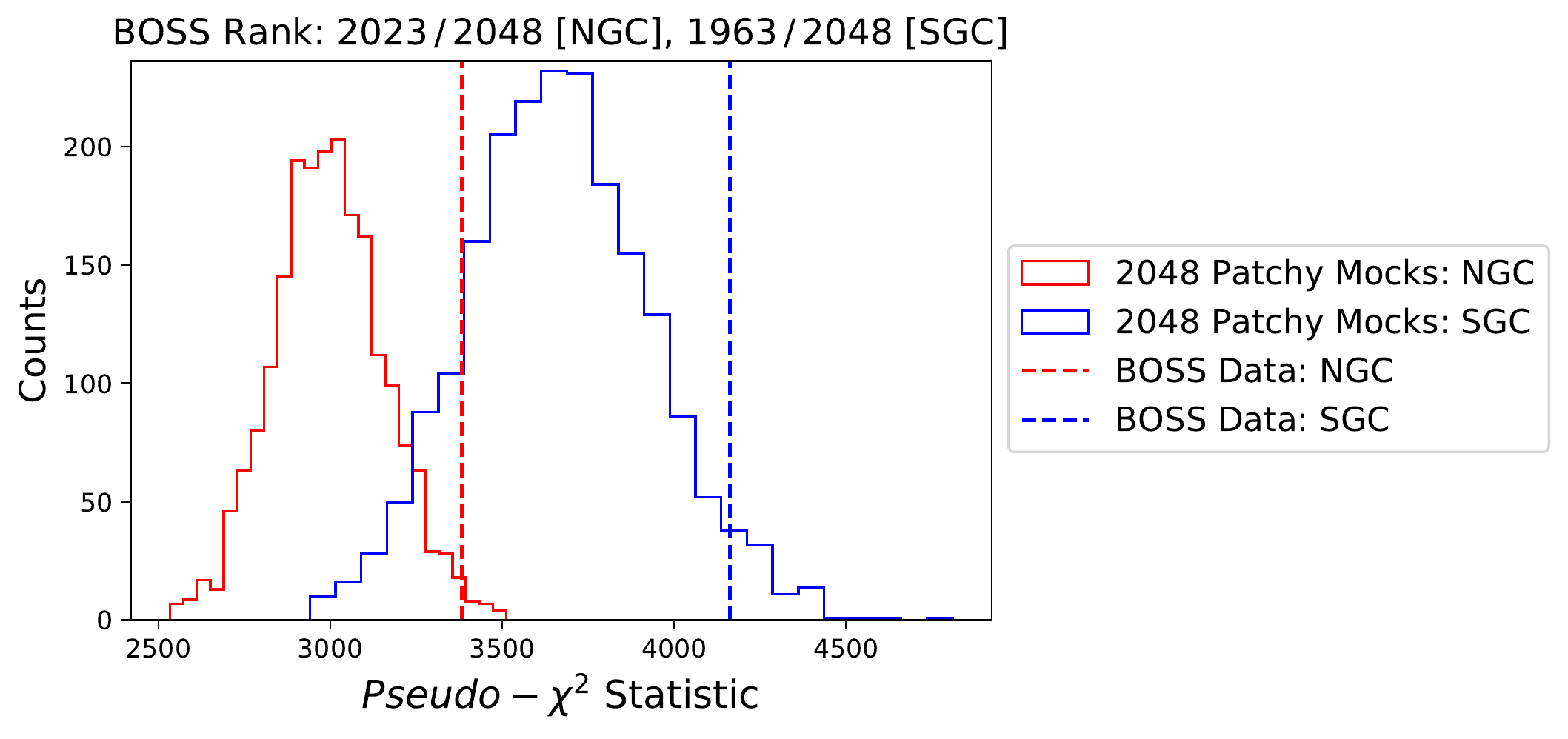}
  \caption{As Fig.\,\ref{fig: rank-test}, but analyzing the two observational regions (NGC and SGC) separately. The NGC region (red) contains almost three times as many galaxies, and is thus expected to be more constraining. We report detection significances of $98.8\%$ ($95.8\%$) for NGC (SGC), corresponding to $2.5\sigma$ ($2.0\sigma$).}
  \label{fig: rank-test-ngc-sgc}
\end{minipage}%
\hfill
\begin{minipage}{.43\textwidth}
  \centering
  \includegraphics[width=0.9\linewidth]{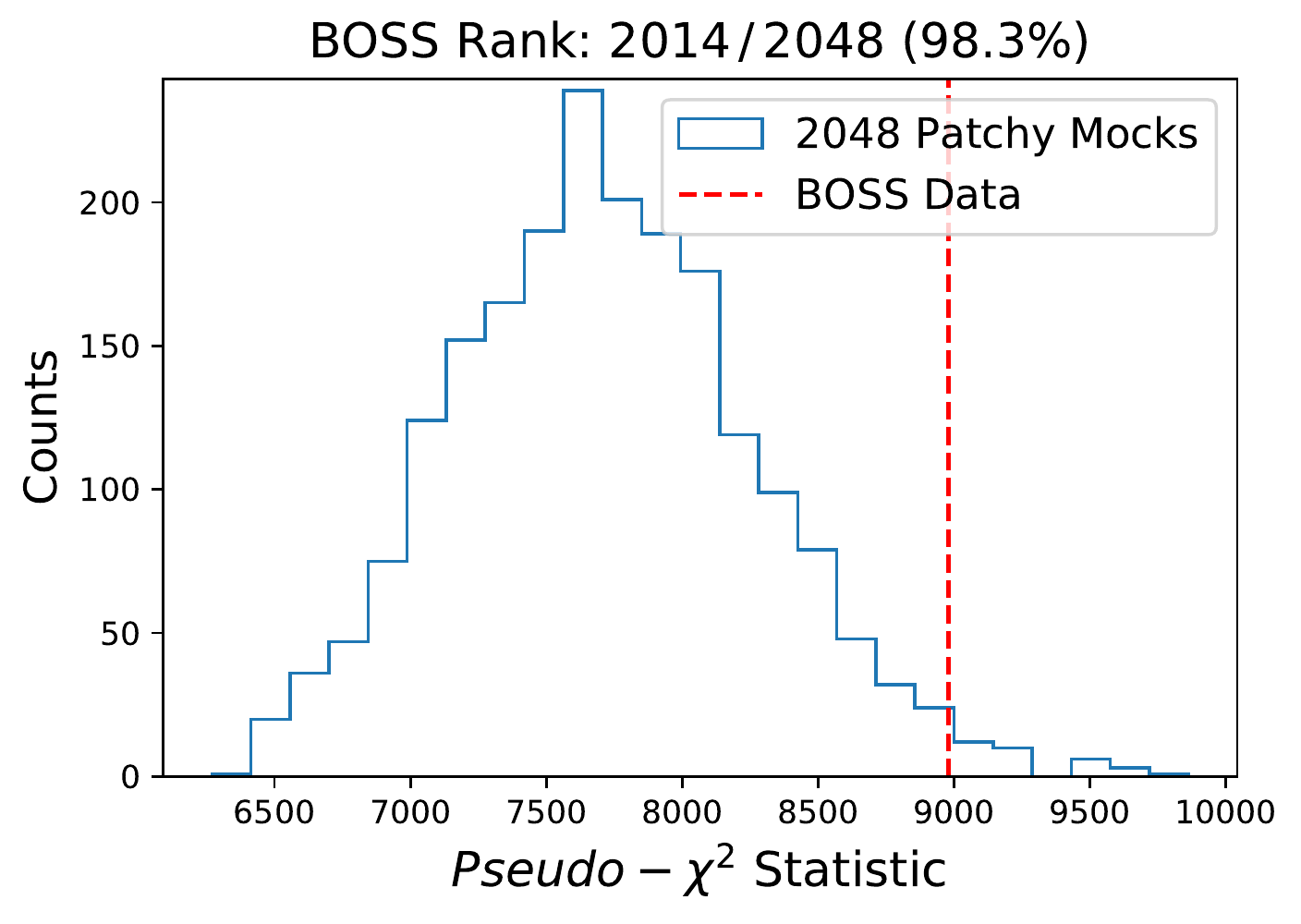}
  \caption{As Fig.\,\ref{fig: rank-test}, but normalizing the 4PCF measurements by a realization-dependent amplitude, as described in \S\ref{subsubsec: sys-rescaling}. This will account for an overall rescaling factor between the BOSS data and \textsc{Patchy} simulations, arising, for example, from a different value of $\sigma_8$. In this case, the detection significance is slightly reduced, however, we still find hints of parity-violation at the $2.4\sigma$ level.}
    \label{fig: rank-test-rescaled}
\end{minipage}
\end{figure}

Fig.\,\ref{fig: rank-test-ngc-sgc} displays the \textit{pseudo}-$\chi^2$ statistic for the two BOSS regions separately (plotting both the observational data and the empirical distribution from 2048 \textsc{Patchy} mocks). Here, we find ranks of $2023/2048$ and $1963/2048$ for the NGC and SGC regions separately, with a lower significance found for the smaller-volume region, as expected. Neither value is enough to claim a detection on its own, however, the trends from the two are in agreement, and their combination reaches the $2.9\sigma$ level reported in \S\ref{sec: data}. This test does not reveal any clear differences between the two regions; the results are consistent with our expectations if this is indeed a \textit{bona fide} detection of parity violation.

\subsection{Dependence on Radial and Angular Scales}\label{subsec: sys-scales}
An additional test is to examine the impact of radial and angular binning on the detection significance reported in \S\ref{sec: results}. For this purpose we will primarily use the rank test of \S\ref{subsec: rank-test}, and adjust the criteria for which bins are included in the analysis, always including a subset of those discussed in \S\ref{subsec: data-4pcf}. In general, we expect the detection significance to decrease somewhat as the dimensionality (and hence number of measured Fourier-modes) is reduced, relative to the initial size of $N_\zeta = 1288$). This will be particularly apparent if we excise regions in which the signal is strongest. 

First, we consider changing the radial binning strategy. Two variations are possible: we can filter based on the distances of secondary galaxies from the primary or those of the secondary galaxies from each other (cf.\,Fig.\,\ref{fig: 4pcf-cartoon}). To test the former, we will separately remove the first and last radial bin in each dimension, \textit{i.e.}\ that with $r_i\in[20,34]\Mpch$ and $[146,160]\Mpch$ respectively. When the minimum radius is increased, we have a modest decrease in dimensionality (to $N_\zeta = 805$) and a detection rank of $1949/2048$ ($2.0\sigma$). If the maximum radius is instead reduced, $N_\zeta$ decreases by a third, but we do not find a change in the overall detection significance. The first result is unsurprising: the variance scales approximately as $(r_1r_2r_3)^{-2}$, thus, if the signal has support over a range of radii, removing the low-$r$ bins would reduce the detection significance. In the second case, the lack of variation suggests that the detection is not caused by some spurious ultra-large mode (arising from foregrounds, for example), though we caution that the signal-to-noise in these modes is the smallest. 

To vary the allowed distances between secondary galaxies, we may modify the restrictions on the allowed tetrahedral shapes. In particular, accepting configurations with $r_1<r_2+\gamma \Delta r<r_3+2\gamma\Delta r$ restricts the internal separations to be at least $\gamma \Delta r$ (with $\gamma=1$ in the fiducial analysis of \S\ref{sec: results}). If we set $\gamma = 0$, and thus include all bins satisfying $r_1<r_2<r_3$, secondary galaxies can be arbitrarily close together, and the size of the data-vector increases to $N_\zeta = 2760$. If the signal contains small-scale power, this should increase the detection significance; here, the non-parametric test gives a rank of \unblind{$2045/2048$} (almost at the saturation point), equivalent to $3.2\sigma$, with the $N_{\rm eig} = 100$ ($250$) compressed $\chi^2$-based analysis giving a detection probability of \unblind{$99.7\%$ ($100.0\%$), equivalent to $3.4\sigma$ ($4.3\sigma$).}
This suggests that small-scale modes are of importance, though we note that they could contain contributions from parity-breaking physics on halo scales, such as magnetohydrodynamics (whose contributions are generally expected to be small in the main analysis, since we restrict to separations above $14\Mpch$). If we instead fix $\gamma = 2$, forcing secondary galaxies to be separated by at least $28\Mpch$, we find a sharp reduction in dimensionality to $N_\zeta = 460$, but only slight decrease in the detection rank (to $1970/2048$, or $2.1\sigma$). This suggests that the signal causing the detection contains non-trivial support on relatively large (and generally well understood) scales.

Next, we consider varying the angular binning, by altering the maximum multipole $\ell_{\rm max}$, and thus the values of $\{\ell_1,\ell_2,\ell_3\}$ (\S\ref{sec: estimator}) that are used in the analysis. Setting $\ell_{\rm max}=3$ reduces the dimensionality to $N_\zeta = 616$, and changes the detection significance to $1970/2048$, or $96.2\%$ ($2.1\sigma$). The physical action of this is to reduce the number of squeezed tetrahedra in the analysis (which have smaller internal tetrahedron angles \textit{i.e.}\ large $\ell$). If $\ell_{\rm max}$ is further reduced to $2$ (with $N_\zeta = 224$) the rank falls to $1466/2048$, or $71.6\%$ ($1.1\sigma$). Our conclusions for $\ell_{\rm max}=3$ are similar to the above: the signal contains contributions from small-scale modes, but is not entirely dominated by them, whilst the $\ell_{\rm max}=2$ result will occur primarily due to the much reduced dimensionality, and thus signal-to-noise. We further note that the multiplets with greatest impact on the detection significance are $\{\ell_1,\ell_2,\ell_3\} = \{1,1,1\}$, $\{1,2,2\}$ and permutations thereof. This is again unsurprising, given that these are the modes with the greatest signal-to-noise, but may serve to indicate that any signal observed is not entirely from some squeezed limit. 

\subsection{Realization-Dependent Rescaling}\label{subsubsec: sys-rescaling}
Under null assumptions (and assuming Gaussianity for the sake of illustration), the rank test compares the variance of the observational data with that of the \textsc{Patchy} mocks. As such, a spurious detection of parity-violation could be caused by the simulations underestimating the true covariance, for example due to mismatches in the galaxy bias parameters or the underlying cosmology. From Fig.\,\ref{fig: rank-test}, we see that excellent agreement between data and mocks can be obtained if one inflates the covariance of the \textsc{Patchy} simulations by $13\%$. Under the assumption of a Gaussian 4PCF covariance, this requires the 2PCF to be rescaled by $3\%$. Whilst this may seem straightforward, it is somewhat non-trivial, since the \textsc{Patchy} simulations are calibrated to match the BOSS two- and three-point clustering statistics on small scales \citep{2016MNRAS.456.4156K,2016MNRAS.460.1173R}. 

To probe this, we consider normalizing the odd-parity 4PCF measurements by their even-parity (more specifically, disconnected) counterparts and reapplying the rank test of \S\ref{subsec: rank-test}. To perform this robustly, we first fit the Gaussian 4PCF contribution to a simple theory template (presented in \citep{2021arXiv210801670P}), extracting a single overall amplitude for each realization.\footnote{An alternative approach is to divide the odd-parity 4PCF by the value of the disconnected (\textit{i.e.}\ Gaussian) 4PCF in each bin; this is not performed since the latter statistic suffers from significant cosmic variance on large scales and can be negative.} The odd-parity 4PCF measurements (both for the BOSS data and each \textsc{Patchy} mock) are then divided by this amplitude, which will remove the dominant effect of an overall rescaling of the \textsc{Patchy} covariance compared to that of BOSS. Whilst the \textsc{Patchy} covariance need not be wrong by a simple constant, this is expected to capture the leading-order effect, and particularly accounts for the unknown value of $b_1\sigma_8$. We note that this prescription cannot be easily applied to the compressed Gaussian analyses, since it violates the Gaussian assumption on large scales due to sample-variance cancellation.

In practice, we find an NGC (SGC) normalization factor of $0.85$ ($0.99$) for BOSS, and $0.92\pm 0.04$ ($0.94\pm0.06$) for the 2048 \textsc{Patchy} simulations. Notably, the BOSS NGC data has a smaller factor; this will shift the corresponding \textit{pseudo}-$\chi^2$ value of BOSS towards those of the mocks. In Fig.\,\ref{fig: rank-test-rescaled}, we show the corresponding rank test results, finding that the detection significance is slightly reduced (to $2014/2048$), matching our expectation. However, this is consistent with the broadened posterior associated with this test (comparing the widths of the empirical distributions in Figs.\,\ref{fig: rank-test}\,\&\,\ref{fig: rank-test-rescaled}), and we still find a weak detection of parity-violation, now at $2.4\sigma$. It is clear that the detection cannot be entirely removed by a simple rescaling; if differences in the statistical properties of simulations and observational data are to blame, they must be scale dependent.

\subsection{Choice of Compression Scheme}\label{subsec: sys-compression}

For the compressed Gaussian analysis of \S\ref{subsec: compressed-analysis}, it is important to choose a set of basis vectors that allow for significant dimensionality reduction whilst preserving key features of the data. For null tests such as that of \S\ref{subsec: compressed-analysis}, this is non-trivial, since we do not have prior knowledge of the form of a parity-violating model. For this reason, the analysis of \S\ref{sec: data} adopted a set of basis vectors selected using a minimum variance criterion; this is equivalent to maximizing signal-to-noise assuming a uniform signal in all bins.

An alternative approach would be to assert some typical form for the parity-violating signal, and use this to select a sensible set of basis vectors onto which the measured 4PCFs are projected. A simple choice is $\zeta_\Lambda(r_1,r_2,r_3)\propto (r_1r_2r_3)^{-1}$ (matching the approximate scaling of the error-bars). In this case, the compressed Gaussian analysis gives detection significances of $0.8\sigma$, $1.8\sigma$, $3.5\sigma$ for $N_{\rm eig} = 50$, $100$, $250$ respectively. A more nuanced choice would be to weight by the inflationary parity-breaking 4PCF model introduced in \S\ref{sec: inflationary-model}. This has a physically reasonable form (albeit specific to a single parity-breaking phenomenon) and leads to Gaussian significances of $0.7\sigma$, $1.5\sigma$, $2.3\sigma$ respectively. Finally, we consider fixing the fiducial model to the $\{\ell_1,\ell_2,\ell_3\} = \{0,0,0\}$ multiplet of the disconnected 4PCF. This will indicate whether the observed signal was sourced by incomplete subtraction of the disconnected component. In this case, we find detection significances of $2.0\sigma$, $1.3\sigma$ and $2.6\sigma$.

For an ideal projection, corresponding to matched true and fiducial 4PCFs, a strong detection significance would be found at small $N_{\rm eig}$ (whereupon the effects of likelihood non-Gaussianity are suppressed), which would increase only slightly as $N_{\rm eig}$ was increased. This behavior is clearly not observed for any of the templates discussed above, suggesting that the signal causing the 4PCF detection is not close to one of the above forms. This is unsurprising: even if the signal is cosmological in nature, it could be sourced by a wide variety of physical effects, each of which has a different signature in the $N_{\zeta} = 1288$-dimensional 4PCF statistic. That we do not observe a strong detection with any template is additionally consistent with the notion that, if such effect is real, it lies on the threshold of that detectable by current data. 

\subsection{Tests on Mock Catalogs}\label{subsec: nseries-testing}
A useful end-to-end test of our analysis pipeline is to apply it to a set of (parity-conserving) mock catalogs and test whether a signal is observed. For this purpose, we will use the \textsc{Nseries} simulation suite:\footnote{Available at \href{https://www.ub.edu/bispectrum/page11.html}{www.ub.edu/bispectrum/page11.html}} a set of 84 mock catalogs created to verify the BOSS analysis pipeline \citep{2017MNRAS.470.2617A}. These were constructed from full $N$-body simulations (though are not quite independent) using the cosmological parameters $\{\Omega_m = 0.286, \sigma_8 = 0.82, n_s = 0.97, h = 0.7, \sum m_\nu = 0\}$, and have similar halo occupation distribution and selection function to the BOSS data, as well as a careful treatment of fiber collisions. The \textsc{Nseries} window function is somewhat different to that of BOSS (cf.\,\S\ref{subsec: data}), and includes only the NGC region. For this reason, the simulations are accompanied by 2048 \textsc{Patchy} simulations generated with the \textsc{Nseries} window function (hereafter `\textsc{Patchy-Nseries}' simulations), which will be used to generate covariances and perform the relevant rank tests. Unlike for BOSS, the \textsc{Patchy-Nseries} simulations were not calibrated to the small-scale clustering of the \textsc{Nseries} simulations (which have a slightly modified cosmology to \textsc{Patchy-Nseries}); as a result, the covariance of \textsc{Nseries} and \textsc{Patchy-Nseries} could differ somewhat, which would have implications for the parity-violation tests, as discussed in \S\ref{subsubsec: sys-rescaling}.

\begin{figure}
    \centering
    \includegraphics[width=0.9\textwidth]{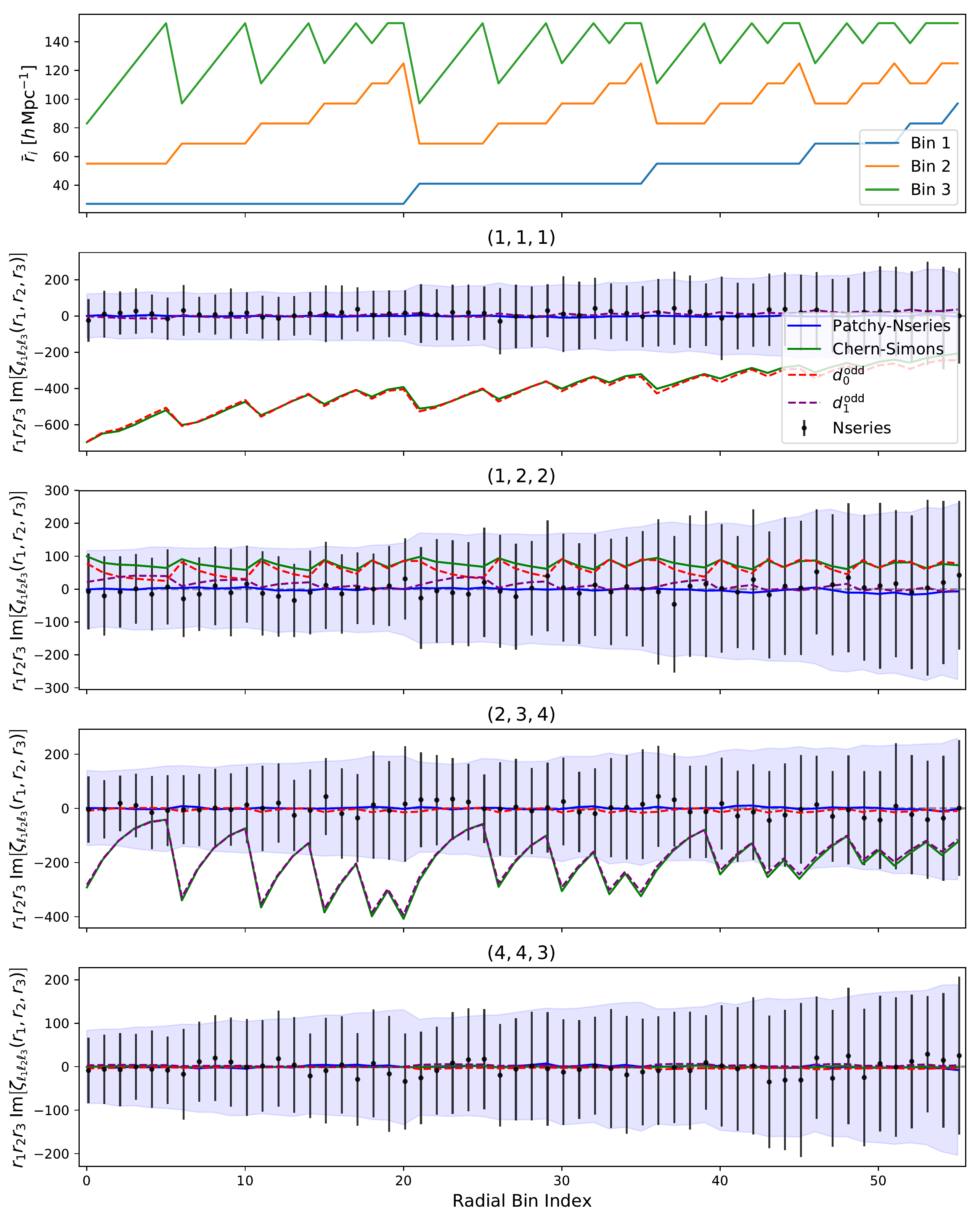}
    \caption{Measurements of the parity-odd 4PCF from the mean of 84 \textsc{Nseries} mocks (black) and 2048 \textsc{Nseries-Patchy} simulations (blue), following the form of Fig.\,\ref{fig: 4pcf-results}. We additionally display theoretical predictions for the Chern-Simons inflationary model (green), and its two constituent parts, proportional to $d_0^{\rm odd}$ (red dashed) and $d_1^{\rm odd}$ (purple dashed). Theory models are multiplied by an amplitude corresponding to $A_{\rm CS} = 5\times 10^4$ for visibility, with data-driven constraints on this parameter presented in Fig.\,\ref{fig: CS-constraints}. The \textsc{Nseries} dataset appears consistent with parity-conservation, as expected; this is explored further in Fig.\,\ref{fig: rank-test-nseries}.}
    \label{fig: 4pcf-results-nseries}
\end{figure}

\begin{figure}
    \centering
    \subfloat[Fiducial\label{fig: rank-test-nseries-fid}]{{\includegraphics[width=0.45\textwidth]{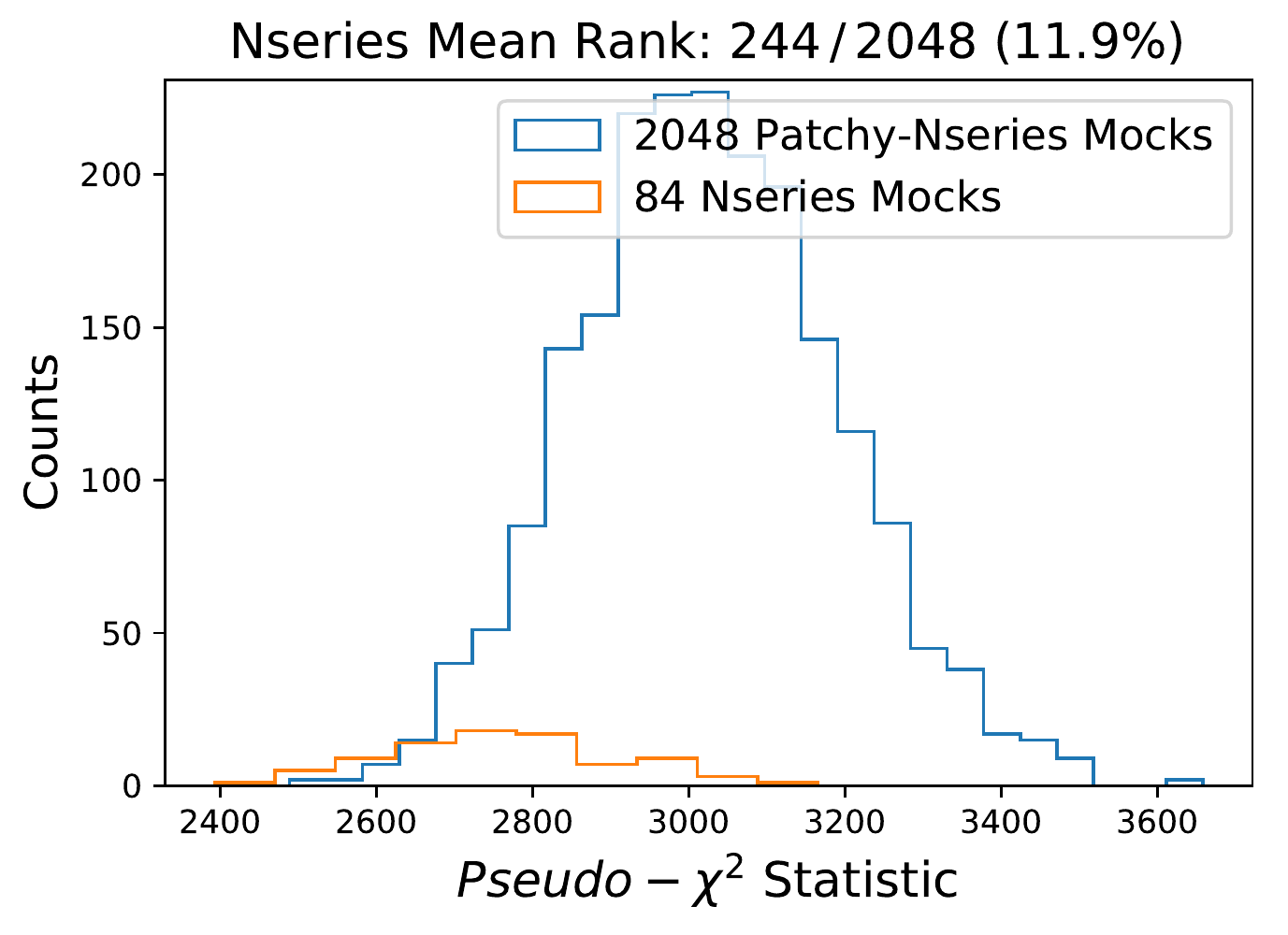}}}%
    \qquad
    \subfloat[Including realization-dependent rescaling\label{fig: rank-test-nseries-rescaled}]{{\includegraphics[width=0.45\textwidth]{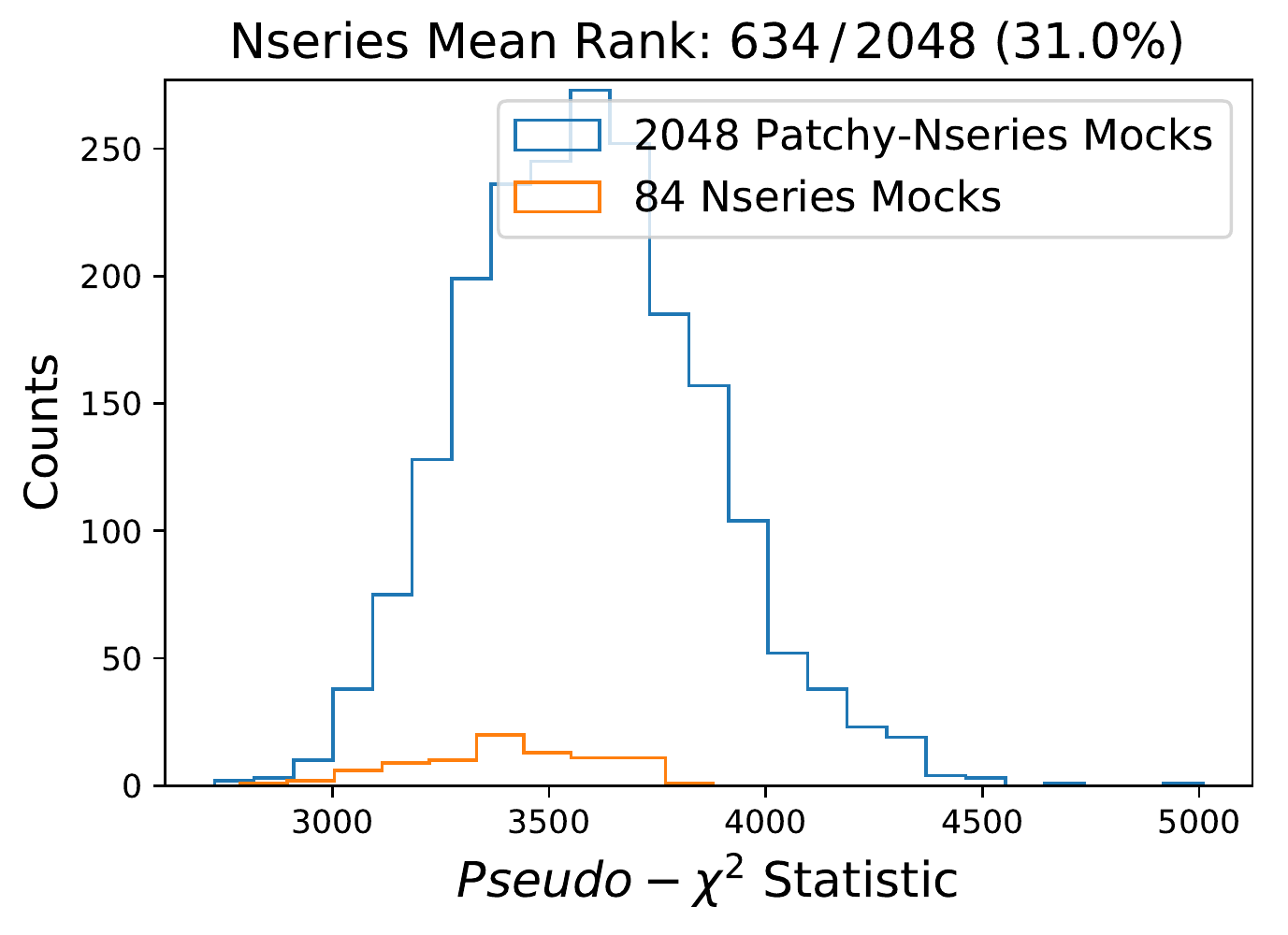}}}
    \caption{Distribution of the \textit{pseudo}-$\chi^2$ statistic for a set of high-fidelity \textsc{Nseries} simulations (orange) and the corresponding empirical distributions obtained from 2048 \textsc{Patchy-Nseries} mocks. In the left panel (analogous to Fig.\,\ref{fig: rank-test}), we find a mean rank of $244/2048$, with $30/84$ mocks lying in the outer $5\%$ region of the empirical distribution \textsc{Patchy-Nseries} distribution. As such, we report an \textit{underestimate} of the test statistic relative to that expected. This is attributed to the difference in cosmology and bias parameters between the \textsc{Nseries} mocks and \textsc{Patchy} simulations. The right panel shows the effect of including the realization-dependent rescaling factor discussed in \S\ref{subsubsec: sys-rescaling}: this reduces the difference between datasets and increases the mean rank to $634/2048$, with only 8 of the 84 simulations in the outer $5\%$ of the empirical distribution. In both cases, there are no \textsc{Nseries} simulations with a \textit{pseudo}-$\chi^2$ value in the upper $5\%$ region of the empirical distribution, thus no detection of parity-violation.}
    \label{fig: rank-test-nseries}
\end{figure}

Using the methodology of \S\ref{sec: analysis-methods}, we analyze each of the 84 \textsc{Nseries} simulations in turn, giving the 4PCF measurements shown in Fig.\,\ref{fig: 4pcf-results-nseries} (which display no obvious signal). Here, we show results only for the rank tests; implementation of the other tests discussed above (such as the compressed Gaussian analysis and constraints on the inflationary models of \S\ref{sec: inflationary-model}) can be found online.\footnote{\href{https://github.com/oliverphilcox/Parity-Odd-4PCF}{github.com/oliverphilcox/Parity-Odd-4PCF}} Fig.\,\ref{fig: rank-test-nseries-fid} compares the distribution of the \textit{pseudo}-$\chi^2$ statistic for the \textsc{Nseries} and \textsc{Patchy-Nseries} simulations (following \S\ref{sec: results}). If our analysis pipeline is working as expected, and the statistical properties of \textsc{Nseries} and \textsc{Patchy-Nseries} are similar, the empirical \textit{pseudo}-$\chi^2$ distributions of the two should match, thus the \textsc{Nseries} simulations should have a mean rank of $\approx 1024/2048$ (with some deviations expected from sample variance). In practice, we find a mean rank of $244/2048$ ($11.9\%$), which is significantly below the expected value. Since the test statistic is a quadratic form, this does not imply a false detection of parity violation (which would give a mean rank \textit{greater} than $1024/2048$); instead it highlights differences between the \textsc{Nseries} and \textsc{Patchy-Nseries} simulation suites (despite the similarity observed in Fig.\,\ref{fig: 4pcf-results-nseries}). As mentioned above, these could come from a variety of factors, such as a different underlying cosmology, the inclusion of fiber collisions in the former, and a different halo occupation distribution. Notably, none of the \textsc{Nseries} mocks have a mean rank \textit{above} $95\%$, hence we do not find evidence for parity-violation in any case.

To more closely assess the impact of differences in the \textsc{Nseries} and \textsc{Patchy-Nseries} covariance matrices (which cause the difference in \textit{pseudo}-$\chi^2$, in the Gaussian limit), we additionally perform a rank test with the mocks rescaled using the realization-dependent factor of \S\ref{subsubsec: sys-rescaling}. The relevant normalization is $0.90\pm0.03$ for \textsc{Nseries}, and $0.92\pm0.04$ for \textsc{Patchy-Nseries}. This differences implies that the \textsc{Nseries} histogram will shift slightly in the direction of \textsc{Patchy-Nseries}; this is observed in Fig.\,\ref{fig: rank-test-nseries-rescaled}. The mean rank is $634/2048$ ($31.0\%$), which is significantly closer to the expected value than without rescaling. Around $10\%$ of the \textsc{Nseries} mocks lie in the outer $5\%$ region of the \textsc{Patchy-Nseries} histogram, giving approximately a one-in-ten chance of a false detection (or here, an anti-detection). The shift in the mean rank induced by the rescaling factor leads to two conclusions: (a) the factor can be usefully adopted to remove the lowest-order differences in simulation covariance matrices (validating the approach of \S\ref{subsubsec: sys-rescaling}), (b) the difference between \textsc{Nseries} and \textsc{Patchy-Nseries} covariances is not fully captured by the rescaling, and is thus scale-dependent.

The above tests indicate that the validity of our null-tests depend strongly on whether the statistical properties of the simulation suite employed to compute empirical distributions match those of the real data. A difference in the sample covariance matrix (which could be caused by various effects, as elaborated upon below) can lead to a false under- or over-detection of parity-violation. For the \textsc{Nseries} simulation suite, these differences are quite significant and lead to a marked under-detection; however, this can be substantially reduced by including the realization-dependent rescaling factor of \S\ref{subsubsec: sys-rescaling}. 

\subsection{Other Sources of Systematics}

Whilst the above tests constrain a variety of different systematic effects, they are by no means exhaustive. Below, we discuss a number of additional effects that could contribute to the tentative detections reported in \S\ref{sec: results}. These can be separated into two groups: (a) effects that lead to the \textsc{Patchy} simulations having a different covariance (and higher-order statistical properties) to that of the data, and (b) observational phenomena causing a parity-violating signal in the odd-parity 4PCF itself.

Firstly, we consider possible causes for a systematic difference between \textsc{Patchy} mocks and BOSS data. Besides simple differences in cosmological parameters or halo occupation distributions, discrepancies could arise due to the treatment of non-linear evolution. The \textsc{Patchy} mocks are generated using approximate gravity solvers only (before being calibrated to an $N$-body simulation, and the observational data); whilst this will not affect the large (linear) scales, it can modify the small-scale clustering, and thus provide an error on small scales, where the constraining power of the data is greatest. Simply calibrating the two- and three-point functions does not guarantee that the covariance is well reproduced, since this depends on $N$-point functions up to $N=8$. Such effects would distort the \textit{pseudo}-$\chi^2$ distribution of the \textsc{Patchy} mocks, and could lead to false detections of parity violation. This hypothesis can be probed via the \textsc{Nseries} tests of \S\ref{subsec: nseries-testing}; once realization-dependent rescaling is included, we find fair agreement between the mock data and \textsc{Patchy-Nseries} simulations, and no spurious detections of parity-violation. For the BOSS data, we expect better agreement with \textsc{Patchy}, given that the simulations were calibrated to the observed small-scale clustering; however, we still find a $2.9\sigma$ detection of parity-violation, or $2.4\sigma$ when the rescaling factor is included. This reduces the likelihood that the effect is caused by inaccuracies in \textsc{Patchy} (considering also the scale-cut information discussed in \S\ref{subsec: sys-scales}), though this effect is certainly worthy of future study.

Another effect not included in the \textsc{Patchy} mocks is that of fiber collisions, arising from the inability to position telescope fibers within a certain distance from each other. If the assignment of fibers to a telescope image is performed in a particular direction (\textit{i.e.}\ picking all objects above a certain brightness from one side of a field-of-view to another), and the quartet of galaxies contains some distant component, a parity-violating signal may be created, due to a preference of one tetrahedral handedness over the other. Two lines of evidence suggest this may not be an important contribution: firstly, the \textsc{Nseries} mocks include the effect of fiber collisions and do not show strong evidence for a parity-violating signal; secondly, most tetrahedra considered in this work have large radial separations. Although fiber collisions happen in the angular domain, rather than radial, all galaxies have minimum separations above $20\Mpch$ (\S\ref{subsec: sys-scales}), thus most fibers of relevance will be spaced by tens of arcminutes. The effect could also change the statistical properties of the data (\textit{i.e.}\ the covariance matrix); this impact of this is analogous to the effects discussed above.

The BOSS data-set is also known to contain systematic effects on large scales due to imperfectly subtracted foreground modes, arising, for example, from Galactic emission or varying seeing conditions. Given that these do not impact BAO measurements, they are usually ignored, though they may be of more relevance to the analyses considered herein. For such an effect to show up in the odd-parity 4PCF signal, it would need to be parity-violating. Since observational effects are not required to obey isotropy and homogeneity, this is possible, and could formed, for example, from the composition of two small-scale 2PCFs (of different lengths) with a large-scale gradient between them. This is analogous to the disconnected 4PCF contribution (but parity-odd) and could be sourced by the above observational phenomena, or even cosmological effects such as isocurvature modes. From the analysis employing hemispherical cuts (\S\ref{subsubsec: sys-ngc-sgc}), it is clear that, if this was the cause of the parity-breaking detection, it is not a one-off phenomenon, and, moreover, it is not sourced solely by very small or very large tetrahedron configurations (\S\ref{subsec: sys-scales}). The particular form of our tetrahedral basis (\textit{i.e.}\ the decomposition into coupled spherical harmonics) makes checking individual tetrahedral configurations difficult, unless a dedicated analysis is performed; the best way to probe them would be with mocks including all observational effects, though none of this type currently exist. We note however, that such large-scale modes would likely have an impact also on lower-order statistics which utilize large-scale data. The consistency of such two-point and three-point function analyses with those of the CMB suggest that these effects, if present, are comparatively small \citep{2020JCAP...05..042I,2021arXiv211204515P}.


\section{Constraints on Inflationary Parity-Violation}\label{sec: inflationary-model}
The 4PCF measurements presented in \S\ref{sec: results} may be used to place constraints on specific models of cosmological parity-violation, such as those involving inflation. As noted in Appendix B of \citep{2020JHEP...04..189L}, parity-violation cannot be generated by single-field inflation,\footnote{\resub{An exception can occur for ghost inflation \citep{Arkani-Hamed:2003juy}; this will be discussed in future work.}} thus its detection in data could give evidence for multiple fields active in the early Universe. Here, we consider a particular multi-field model, which gives an analytic form for the parity-odd galaxy 4PCF, in addition to lower-dimensional observables. An analogous procedure may be used to constrain any model which induces a non-trivial parity-odd 4PCF; a selection of these are briefly discussed in \S\ref{subsubsec: parity-breaking-sources}.

\subsection{Primordial Correlators}
\subsubsection{Chern-Simons Inflation}
Consider an inflationary Lagrangian containing the following couplings between an inflaton field, $\phi$, and a $U(1)$ gauge field, $A_\mu$:
\beq\label{eq: lag-cs}
    \mathcal{L}\supset f(\phi)\left[-\frac{1}{4}F_{\mu\nu}F^{\mu\nu}+\frac{\gamma}{4} F_{\mu\nu}\tilde F^{\mu\nu}\right],
\eeq
where $F_{\mu\nu}\equiv \partial_\mu A_\nu - \partial_\nu A_\mu$ is a two-form. The second term involves the Hodge dual, $\tilde F^{\mu\nu}$, and is of the Chern-Simons form \citep[e.g.,][]{1990PhRvL..65.3233F,2015JCAP...07..039B,2017JCAP...07..034B,2017JCAP...07..034B,2013JCAP...04..046A}. This is controlled by two pieces: a function $f(\phi)$ giving the time evolution of the field, and a dimensionless ratio $\gamma$, which sets the amplitude of parity-breaking.\footnote{Constraints on the reheating temperature from Big Bang nucleosynthesis restrict the coupling strength to $|\gamma|<5.5$ \citep{2004PhRvD..70d3506H,2015JCAP...01..027B}.} Following \citep{2014JCAP...10..056C,2015JCAP...07..039B}, we will assume $\gamma$ to be constant (on naturalness grounds), and fix $f(\phi)\propto a^{-4}$ (for scale-factor $a$), such that the vector field has a time-independent vacuum expectation value (vev) and thus a scale-invariant correlator.\footnote{This is a natural choice within the Ratra mechanism \citep{1992ApJ...391L...1R,2014JCAP...10..056C}.} If one instead assumes $f(\phi) \propto \mathrm{const.}$, the energy density of the vector field will decay as $a^{-4}$ during inflation, giving observational signatures only on the largest scales \citep{2015JCAP...01..027B}. 

The Lagrangian \eqref{eq: lag-cs} leads to a number of modifications to the standard inflationary picture. Firstly, the presence of a background vev $A^\mu_0$ (usually represented in the electromagnetic notation as $\vec E_0$, with $\vec B_0 = \vec 0$) leads to anisotropic expansion, and its perturbations can provide an isocurvature source for the primordial curvature $\zeta$. Such effects are strongly constrained by CMB data, limiting the energy density in the gauge field (hereafter $\rho_E$) to be a small fraction of the inflaton energy density $\rho_\phi$. Couplings between the inflaton and gauge field will additionally generate gravitational waves through the metric tensor $h_{ij}$, as well as scalar-tensor couplings. This can lead to observable signatures in CMB $E$-modes and $B$-modes (and non-zero parity-breaking spectra such as $C_\ell^{TB}$ and $C_\ell^{EB}$), though such effects are slow-roll suppressed \citep{2015JCAP...07..039B}. We note that a non-zero vev $\vec E_0$ is a natural prediction of the theory; this is simply the impact of long-wavelength classical perturbations in the vector field which have not yet re-entered the horizon.\footnote{Following \citep{2015JCAP...07..039B}, the magnitude of this ought to scale as $N_{\rm tot}-N$, where $N$ is the number of $e$-folds before the CMB modes exited the horizon, and $N_{\rm tot}$ gives the total duration of inflation.} Furthermore, the action of non-zero $\gamma$ is to produce an excess of one gauge field helicity mode over the other, causing a parity asymmetry.

In this work, we are interested in the scalar correlators generated by the above interaction, \textit{i.e.}\ those only involving the curvature $\zeta$. As demonstrated in \citep{2015JCAP...01..027B,2015JCAP...07..039B,2014JCAP...04..027S,2016PhRvD..94h3503S}, the gauge fields lead to anisotropy in the two-point function:
\beq\label{eq: CS-2pcf}
    \av{\zeta(\vk_1)\zeta(\vk_2)} =  \delD{\vk_1+\vk_2}P_\zeta(k_1)\left[1+g_\ast (\hk_1\cdot\hat{\vec E_0})\right],
\eeq
where $P_\zeta \approx H^2/(4\epsilon M_p^2k^3)$ is the primordial power spectrum, for Hubble parameter $H$, Planck mass $M_p$, and slow-roll parameter $\epsilon$. This is of the well-known ACW form \citep{2007PhRvD..75h3502A}, for approximately scale-invariant coupling $g_\ast\propto \rho_E/\rho_\phi$. Furthermore, the Lagrangian given in \eqref{eq: lag-cs} generates an angle-averaged bispectrum and trispectrum of the curvature perturbation $\zeta$. The first takes the form of \citep{2013JCAP...05..002S,2020A&A...641A...9P}:
\beq\label{eq: CS-3pcf}
    \av{\zeta(\vk_1)\zeta(\vk_2)\zeta(\vk_3)} = \delD{\vk_1+\vk_2+\vk_3}\sum_{n=0,1,2} c_n \mathcal{L}_n(\hk_1\cdot\hk_2)P_\zeta(k_1)P_\zeta(k_2) + \text{2 perms.},
\eeq
where $\mathcal{L}_n$ is a Legendre polynomial of order $n$, and the coefficients $c_n$ (simply related to \textit{Planck}'s $f_{\rm NL}^{L=n}$ parameters \citep{2020A&A...641A...9P}) are again proportional to the fractional energy density in the vector field \citep{2015JCAP...07..039B}, and satisfy $c_0=-2/3 c_1=2c_2$. These parametrize a number of effects beyond the inflationary Lagrangian \eqref{eq: lag-cs}, such as curvature fluctuations sourced by primordial magnetic fields \citep[e.g.,][]{2012JCAP...06..015S}, solid inflation \citep[e.g.,][]{2013JCAP...10..011E} (which boasts $c_2\gg c_0$), and massive spinning particles \citep[e.g.,][]{2015arXiv150308043A,2018PhRvD..98d3533F}. Finally, the four-point function for the CS model was computed in \citep{2014JCAP...04..027S,2016PhRvD..94h3503S}, and can be expressed in terms of the reduced trispectrum $t$, defined as:
\beq
    \av{\prod_{i=1}^4\zeta(\vk_i)} = (2\pi)^3\delta_{\rm D}\left(\sum_{i=1}^4 \vk_i\right)\left[t^{\vk_1\vk_2}_{\vk_3\vk_4} + \text{ 23 perms.}\right].
\eeq
Separating even and odd parts, we have:
\beq\label{eq: tk-even-odd}
    \left.t^{\vk_1\vk_2}_{\vk_3\vk_4}\right|_{\rm even} &=& \sum_{n=0,2} d_n^{\rm even} \left(\mathcal{L}_n(\hk_1\cdot\hk_3)+\mathcal{L}_n(\hk_1\cdot\hat{\vec K})+\mathcal{L}_n(\hk_3\cdot\hat{\vec K})\right)P_\zeta(k_1)P_\zeta(k_3)P_\zeta(K)\\\nonumber
    \left.t^{\vk_1\vk_2}_{\vk_3\vk_4}\right|_{\rm odd} &=& i\sum_{n=0,1} d_n^{\rm odd} \left(\mathcal{L}_n(\hk_1\cdot\hk_3)+\mathcal{L}_n(\hk_1\cdot\hat{\vec K})+(-1)^n\mathcal{L}_n(\hk_3\cdot\hat{\vec K})\right)\left[(\hk_1\times\hk_3)\cdot\hat{\vec K}\right]P_\zeta(k_1)P_\zeta(k_3)P_\zeta(K)
\eeq
where $\vK = \vec k_3+\vec k_4$ is the trispectrum diagonal, and $t_{-\vk_3-\vk_4}^{-\vk_1-\vk_2}=\pm t_{\vk_3\vk_4}^{\vk_1\vk_2}$ for the even and odd components respectively. This depends on new parameters, $\{d_n\}$, with the odd-parity contributions sourced iff $\gamma$ is non-zero (\textit{i.e.}\ if the Lagrangian contains a Chern-Simons term). Whilst \eqref{eq: tk-even-odd} is a relatively general parametrization for scale-invariant parity-odd primordial trispectra, we may specialize to the Chern-Simons model by fixing $d_0^{\rm even} = d_2^{\rm even}/2$, and $d_0^{\rm odd} = -d_1^{\rm odd}/3$, each of which is linear in $\rho_E/\rho_\phi$. 

For later convenience, we will rewrite the odd-parity reduced trispectrum corresponding to \eqref{eq: lag-cs} as
\beq\label{eq: tk-CS}
    \left.t^{\vk_1\vk_2}_{\vk_3\vk_4}\right|_{\rm odd} = 3i\,A_{\rm CS}\,P_\zeta(k_1)P_\zeta(k_3)P_\zeta(s)\left[(\hk_1\times\hk_3)\cdot\hs\right]\left[1-\hk_1\cdot\hk_3+\hk_1\cdot\hs-\hk_3\cdot\hs\right]\equiv t(\vk_1,\vk_3,\vs),
\eeq
introducing the Mandelstam variable $\vs \equiv \vk_1+\vk_2$, and defining the overall amplitude $A_{\rm CS}\equiv-d_0^{\rm odd}=d_1^{\rm odd}/3$ as
\beq\label{eq: A-CS-def}
    A_{\rm CS} \approx \frac{0.3}{\pi^2}\frac{e^{8\pi |\gamma|}}{|\gamma|^6}\frac{|g_*|}{0.01}\left(\frac{N}{60}\right)^2, \qquad g_*\approx -\frac{5.4\times 10^5}{\pi}\frac{e^{4\pi|\gamma|}}{|\gamma|^3}\frac{0.01}{\epsilon}\left(\frac{N}{60}\right)^2\frac{\rho_E}{\rho_\phi},
\eeq
for $|\gamma|>1$, assuming that modes of interest exited the horizon $N$ $e$-folds before the end of inflation. As expected from \S\ref{sec: estimator}, the reduced trispectrum in \eqref{eq: tk-CS} is pure imaginary, and depends on a parity-odd cross-product. 

Combining the above, we can constrain the gauge field energy density from the two-, three- or four-point functions of the scalar field, or, following the transformations outlined in \ref{subsec: CS-model}, the galaxy overdensity. As in \citep{2015JCAP...07..039B,2016PhRvD..94h3503S}, we expect the constraints to be a strong function of $|\gamma|$: gauge field production increases with $\gamma$, and the $N$-point function scales as $2(N-1)$ powers of the gauge field perturbation $\delta E$, leading to a prefactor of $e^{4\pi(N-1)|\gamma|}/|\gamma|^{3(N-1)}$. In this work, we will derive constraints only from the parity-odd 4PCF: this provides both a tight constraint for $|\gamma|>0$ due the above argument, and is a clean probe, since it is not contaminated by gravitational non-Gaussianity.


\subsubsection{Other Sources}\label{subsubsec: parity-breaking-sources}
Before proceeding to derive constraints on the Chern-Simons inflationary model, we briefly discuss a number of alternative phenomena that can source parity violation. Firstly, the Lagrangian presented in \eqref{eq: lag-cs} remains applicable when $\phi$ is \textit{not} the inflaton. This has been suggested as a mechanism for primordial magnetogenesis \citep{2014JCAP...10..056C,1992ApJ...391L...1R}, and will source similar parity violation. An additional case of interest is when $F_{\mu\nu}$ is the electromagnetic tensor and $\phi$ an axion-like particle in the late Universe; this can generate detectable cosmic birefringence \citep[e.g.,][]{2010PhRvD..81l3529G,2020PhRvL.125v1301M,Campeti:2022acx}. Secondly, the inflationary Lagrangian could contain a \textit{gravitational} Chern-Simons term
\beq\label{eq: lag-grav-cs}
    \mathcal{L} \supset f'(\phi)R^\lambda_{\sigma\mu\nu}\tilde R_{\lambda}^{\sigma\mu\nu}
\eeq
\citep{1999PhRvL..83.1506L,2009PhR...480....1A,2017JCAP...07..034B}, for some Riemann curvature tensor $R_{\mu\nu\rho\sigma}$ and a function $f'(\phi)$ of the inflaton $\phi$. Following a similar calculation to that of \citep{2016PhRvD..94h3503S} for the Lagrangian described above, one can compute to compute the scalar trispectrum corresponding to \eqref{eq: lag-grav-cs}, which is now sourced by couplings to tensor modes rather than vectors. 

Primordial parity-violation can also arise from \textit{particle exchange} in the early Universe (as part of the so-called `cosmological collider' \citep{2003JHEP...05..013M,2015arXiv150308043A}). At high energies, it is natural to assume that the inflaton is coupled to additional fields (be they scalars, vectors or tensors), via some three-point interaction vertices, e.g., $(\partial_\mu\phi)(\partial_\nu\phi)\partial^\mu V^\nu$ for a primordial vector $V^\mu$. In this case, the two-point function of the inflaton (and thus the curvature perturbation $\zeta$) \resub{can} receive an off-diagonal contribution. \resub{A simple example of this occurs for light mediators, which takes the form}
\beq
    \left.\av{\zeta(\vk_1)\zeta(\vk_2)}\right|_{X_p(\vK)} = \frac{B_{\zeta\zeta {X_p}}(k_1,k_2,K)}{P_{X_p}(K)}X_p^*(\vK)\delD{\vk_1+\vk_2+\vK},
\eeq
where $P_{X_p}$ is the power spectrum of $X$ in some polarization state $p$, and $B_{\zeta\zeta X_p}$ is the interaction three-point function \resub{\citep[e.g.,][]{2012PhRvL.108y1301J,2013PhRvD..87j3006D,Dimastrogiovanni:2022afr,Dimastrogiovanni:2015pla,2014JCAP...12..050D}}, \resub{or, more precisely, its consistency-relation-violating component \citep{Creminelli:2004yq}}. Combining two of these interactions, one obtains an exchange diagram for $\zeta$, \textit{i.e.}\ a four-point function of the scalar curvature, with some shape proportional to $|B_{\zeta\zeta X_p}|^2/P_{X_p}$. If the field in question is helical, this can lead to a parity-violating trispectrum if $P_{X_+}\neq P_{X_-}$. As discussed in \citep{2012PhRvL.108y1301J}, such contributions can be measured directly using a statistical anisotropy (`fossil') estimator, which can be recast as a model-specific compression of the full four-point function. In the squeezed limit ($K\ll k_1\approx k_2$, \resub{again noting that massive mediators can have different behaviors, \citep[cf.][]{Dimastrogiovanni:2015pla,Dimastrogiovanni:2022afr,2016JCAP...06..052B}}), the interaction bispectrum is approximately given by $(3/2)P_\zeta(k_1)P_{X_p}(K)\epsilon^p_{ij}\hat k_1^i\hat k_1^j$ \citep{2003JHEP...05..013M,2015arXiv150308043A,2012PhRvL.108y1301J}, where $\epsilon^p$ is the polarization tensor. For a \resub{light} vector exchange field $X$, this leads to the following schematic form for the trispectra (following \citep{2012PhRvL.108y1301J}):
\beq
    \left.t^{\vk_1\vk_2}_{\vk_3\vk_4}\right|_{\rm even} &\sim& \left[P_{X_+}(K)+ P_{X_-}(K)\right]P_\zeta(k_1)P_\zeta(k_3)(\hk_1\cdot\hat{\vK})(\hk_3\cdot\hat{\vK})\left[\hk_1\cdot\hk_3-(\hk_1\cdot\hat{\vK})(\hk_3\cdot\hat{\vK})\right],\\\nonumber
    \left.t^{\vk_1\vk_2}_{\vk_3\vk_4}\right|_{\rm odd} &\sim& -i\left[P_{X_+}(K)- P_{X_-}(K)\right]P_\zeta(k_1)P_\zeta(k_3)(\hk_1\cdot\hat{\vK})(\hk_3\cdot\hat{\vK})\left[(\hk_1\times\hk_3)\cdot\hat{\vK}\right]
\eeq
The similarities of this and \eqref{eq: tk-even-odd} are manifest, particularly if one assumes a scale invariant form for the power spectrum of $X$, such that $P_{X_\pm}(K)\propto P_\zeta(K)$. We thus note that a simple extension to the parametrization of \eqref{eq: tk-even-odd} can incorporate trispectra arising from particle exchange. A similar conclusion holds also for intermediate fields $X$ of higher spin (noting that scalar exchange cannot generate parity-violating couplings).



\subsection{4PCF Model}\label{subsec: CS-model}

To place constraints on the Chern-Simons interaction, we must compute the galaxy 4PCF associated with the primordial trispectrum of \eqref{eq: tk-CS}. At redshift $z$, the tree-level galaxy trispectrum, $T_g$, can be related to the primordial correlator via
\beq\label{eq: trispectrum-gal-prim}
    (2\pi)^3\delta_{\rm D}(\vk_1+\vk_2+\vk_3+\vk_4)T_g(\vk_1,\vk_2,\vk_3,\vk_4,z) = \av{\prod_{i=1}^4Z_1(\vk_i,z)M(k_i,z)\zeta(\vk_i)},
\eeq
where $M(k,z)$ is the transfer function, defined by $\delta_{\rm matter}(\vk,z) \equiv M(k,z)\zeta(\vk)$, and $Z_1(\vk,z) \equiv b(z)+f(z)(\hk\cdot\hn)^2$ is the tree-level galaxy RSD kernel (for linear bias $b(z)$, growth-rate $f(z)$ and line-of-sight $\hn$). From \eqref{eq: trispectrum-gal-prim}, we may obtain the odd-parity 4PCF using Fourier transforms:
\beq\label{eq: CS-4PCF-overall}
    \zeta_-(\vr_1,\vr_2,\vr_3,z) &=& \prod_{i=1}^4\left[\int_{\vk_i}M(k_i,z)Z_1(\vk_i,z)\right]\int_{\vs}t(\vk_1,\vk_3,\vs)\left[e^{i(\vk_1\cdot\vr_1+\vk_2\cdot\vr_2+\vk_3\cdot\vr_3)} + \text{ 23 perms.}\right]\\\nonumber
    &&\,\times\,\delD{\vk_1+\vk_2-\vs}\delD{\vk_3+\vk_4+\vs},
\eeq
shifting the permutation sum to the exponential term by symmetry, and using Dirac delta functions to enforce $\vs = \vk_1+\vk_2 = -\vk_3-\vk_4$. The corresponding multiplets, $\zeta_{\ell_1\ell_2\ell_3}$, can then be estimated by performing weighted integrals over $\hr_i$, as in \eqref{eq: 4pcf-basis-coeffs}. Following a lengthy derivation sketched in Appendix \ref{appen: chern-simons}, we obtain the final form for odd $\ell_1+\ell_2+\ell_3$:
\beq\label{eq: zeta-cs-main-text}
    \zeta_{\ell_1\ell_2\ell_3}(r_1,r_2,r_3,z) &=& (4\pi)^{11}\sqrt{2}A_{\rm CS}\,i^{\ell_1+\ell_2+\ell_3}\sum_{L_1L_2L_3L_4L_5L_5'}i^{L_1+L_2+L_3+L_4-L_5+L_5'}\mathcal{C}_{L_1L_2L_3L_4L_5L_5'}\\\nonumber
    &&\,\times\,\tjo{L_1}{L_2}{L_5}\tjo{L_3}{L_4}{L_5'}\mathcal{M}^{\ell_{1}\ell_{2}(\ell_3)\ell_{3}0}_{L_1L_2L_3L_4L_5L_5'}\\\nonumber
    &&\,\times\,\int x^2dx\,\int x'^2dx'\,K_{L_5L_5'}(x,x')I_{L_1}^{\ell_1}(x;r_{1})J_{L_2}^{\ell_{2}}(x;r_{2})I_{L_3}^{\ell_{3}}(x';r_{3})J_{L_4}^{0}(x';0) + \text{23 perms.}
\eeq
(cf.\,\ref{eq: zeta-cs-appendix}). Here, $I$, $J$ and $K$ are Bessel-weighted integrals over the transfer function and/or primordial power spectrum \eqref{eq: I,J,K-def}, $\mathcal{C}_{\ell_1\cdots\ell_N} \equiv \sqrt{(2\ell_1+1)\cdots(2\ell_N+1)}$, and $\mathcal{M}$ is a coupling matrix given in \eqref{eq: coupling-matrix}. The 4PCF may thus be computed as a two-dimensional integral, following evaluation of the (radially-binned) $I$, $J$ and $K$ functions for a range of values of $L$, $\ell$, $x$ and $x'$. In practice, the 4PCF model is computed in \textsc{Python}, with the various Wigner 3-$j$ and 9-$j$ symbols evaluated using \textsc{sympy}.\footnote{Code implementing this calculation is available at \href{https://github.com/oliverphilcox/Parity-Odd-4PCF.git}{github.com/oliverphilcox/Parity-Odd-4PCF}.} In Fig.\,\ref{fig: 4pcf-results-nseries} we plot the theoretical model for a range of multiplets, finding a shape that depends strongly on both $\{r_i\}$ and $\{\ell_i\}$, with some multiplets dominated by the $d_0^{\rm odd}$ part, and others by that proportional to $d_1^{\rm odd}$. 

\subsection{Amplitude Constraints}\label{subsec: CS-results}
The importance of the inflationary gauge field may be quantified by the ratio of energy densities, $\rho_E/\rho_\phi$, or the parity-odd 4PCF amplitude, $A_{\rm CS}$ \eqref{eq: A-CS-def}. \unblind{This sets the level of parity-violation imprinted in the primordial inflaton perturbations, arising from interactions with the $U(1)$ gauge field. To constrain the amplitudes}, we perform parameter inference using the measured 4PCF multiplets of \S\ref{sec: results} and the Chern-Simons model given in \eqref{eq: zeta-cs-main-text}. For simplicity, we will assume the data to be Gaussian distributed, and work in a compressed subspace containing $N_{\rm eig} = 100$ basis vectors for each of the NGC and SGC regions (whose distribution was shown to be approximately Gaussian in \S\ref{sec: results}). In \S\ref{subsec: compressed-analysis}, a minimum-variance criterion was used to select the basis vectors; here, we instead pick those with maximal signal-to-noise for the Chern-Simons model. The reduced dimensionality facilitates direct use of the \textsc{Patchy} simulations to form the sample covariance \eqref{eq: compressed-sample-cov}; to account for the finite number of mocks, we perform inference using the following log-likelihood:
\beq
    -\log L(A_{\rm CS}) = \frac{N_{\rm mocks}}{2}\log\left[1+\frac{T^2(A_{\rm CS})}{N_{\rm mocks}-1}\right] + \text{const.}
\eeq
This uses the $T^2$ statistic, defined analogously to \eqref{eq: T2-def}:
\beq
    T^2(A_{\rm CS}) \equiv \left(\hat v_{\rm data}-A_{\rm CS}v_{\rm CS}\right)^T\hat C_v^{-1}\left(\hat v_{\rm data}-A_{\rm CS}v_{\rm CS}\right),
\eeq
where $\hat v_{\rm data}$ represents the compressed 4PCF data-vector, $\hat C_v$ is a sample covariance, and $v_{\rm CS}$ is the compressed 4PCF model of \eqref{eq: zeta-cs-main-text}, excluding the $A_{\rm CS}$ prefactor. Likelihoods for the NGC and SGC region are constructed separately and multiplied, assuming independence. Here, we perform two analyses; one using the BOSS data, and the second using the mean of $N_{\rm mocks} = 2048$ \textsc{Patchy} mocks. In the latter case, no Chern-Simons contribution should be present.

\begin{figure}
    \centering
    \includegraphics[width=0.45\textwidth]{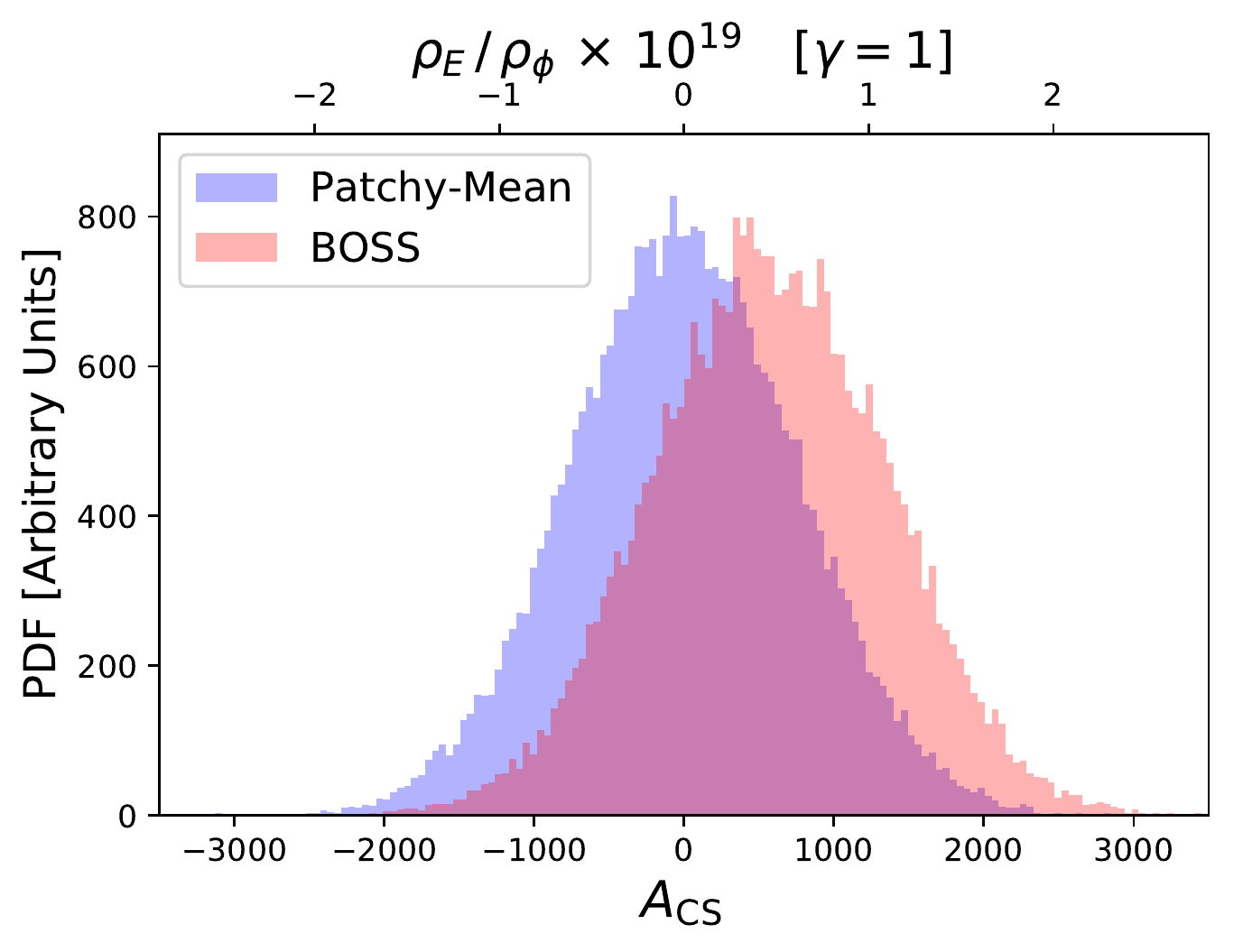}
    \includegraphics[width=0.45\textwidth]{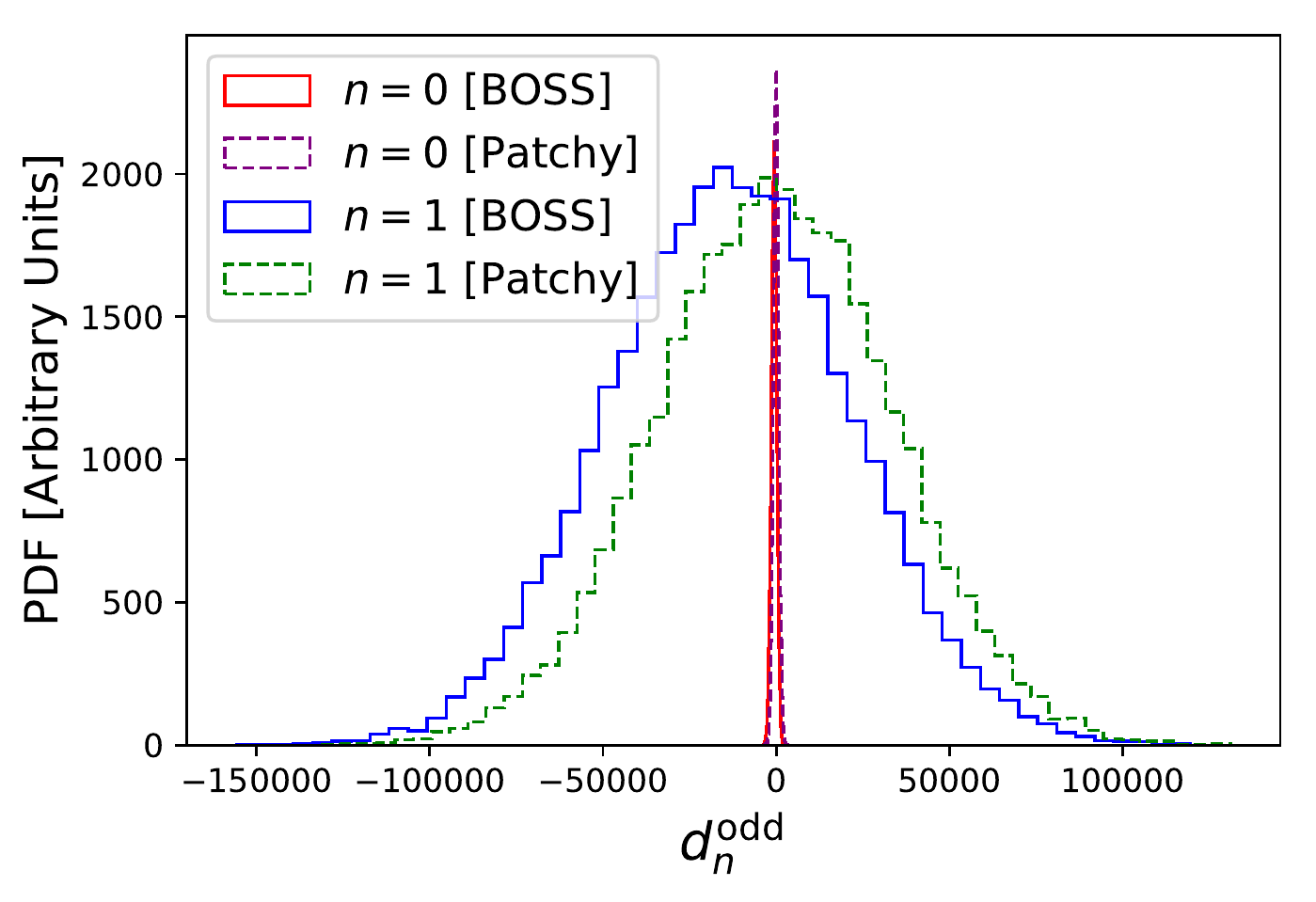}
    \caption{Constraints on the amplitude of physical models for parity-violation, both for the Chern-Simons Lagrangian of \eqref{eq: lag-cs} (left) and the more general parametrization of \eqref{eq: tk-even-odd} (right), using the explicit parity-odd 4PCF prediction given in \eqref{eq: zeta-cs-main-text} and plotted in Fig.\,\ref{fig: 4pcf-results-nseries}. To obtain these distributions, we fit the measured 4PCF multiplets shown in Fig.\,\ref{fig: 4pcf-results} to templates derived in Appendix \ref{appen: chern-simons}, following compression of both observations and model into a 100-dimensional subspace. For the Chern-Simons model, we give results for both the trispectrum amplitude $A_{\rm CS}$ and the corresponding ratio of parity-breaking gauge field and inflaton energy densities, $\rho_E/\rho_{\phi}$, using relation \eqref{eq: A-CS-def} with $|\gamma| = 1$. The $1\sigma$ constraints are
    \unblind{$A_{\rm CS} = 0\pm760$} 
    for the mean of 2048 \textsc{Patchy} mocks (blue) and 
    \unblind{$A_{\rm CS} = 570\pm780$} 
    for the BOSS data (red), both of which are consistent with zero. The right panel gives constraints on the $d_n^{\rm odd}$ parameters appearing in \eqref{eq: tk-even-odd}: we find $d_0^{\rm odd} = -610\pm800$ ($-20\pm770$) and $d_1^{\rm odd} = -13\,000\pm35\,000$ ($0\pm 34\,000$) for BOSS (mean of Patchy) data respectively. In the Chern-Simons model, $-d_0^{\rm odd} = d_1^{\rm odd}/3=A_{\rm CS}$.}
    \label{fig: CS-constraints}
\end{figure}

Fig.\,\ref{fig: CS-constraints} shows the resulting constraints on the trispectrum amplitude. For the mean of \textsc{Patchy} mocks, the $1\sigma$ constraint is 
\unblind{$A_{\rm CS} = 0\pm760$} 
(demonstrating unbiasedness, as expected), with 
\unblind{$A_{\rm CS} = 570\pm780$} 
observed for BOSS. In both cases, \unblind{the constraints are consistent with zero, suggesting that the Chern-Simons coupling is not responsible for the detection of parity-violation reported in} \S\ref{sec: results}. If we additionally restrict to $A_{\rm CS}\geq 0$, we find that
\unblind{$A_{\rm CS}<1500$} and \unblind{$A_{\rm CS}<2000$} 
for the mean-of-\textsc{Patchy} and BOSS datasets respectively (95\% CL).\footnote{We have additionally verified that no false detection of Chern-Simons inflation is obtained when using the \textsc{Nseries} mocks of \S\ref{subsec: nseries-testing}.} Additionally including tetrahedra with small separations between secondary galaxies (as in \S\ref{sec: results}) does not appreciably improve the constraints, which we justify by noting that the bulk of the signal-to-noise occurs on comparatively large scales.

Of greater physical interest are the constraints on the energy densities $\rho_E/\rho_\phi$. These may be obtained from the $A_{\rm CS}$ bounds using \eqref{eq: A-CS-def}, assuming fiducial values for the inflationary parameters and fixing the coupling strength to $\gamma=1$, giving an equal contribution from the parity-even and parity-odd terms in \eqref{eq: lag-cs}.\footnote{Note that exchanging $\gamma\to-\gamma$ simply swaps the dominant and suppressed helicity states of the gauge field.} Using the BOSS CMASS data, we find 
\unblind{$\rho_E/\rho_\phi<1.6\times 10^{-19}$} ($95\%$ CL). If $|\gamma|$ is increased to $2$, the gauge field production is strongly amplified, and the constraint tightens to
\unblind{$\rho_E/\rho_\phi<3.5\times 10^{-33}$}.

We may additionally place limits on the phenomenological parameters $\{d_n^{\rm odd}\}$ appearing in \eqref{eq: tk-even-odd}. Using an analogous method to the above, we find the $1\sigma$ constraints $d_0^{\rm odd} = -610\pm 800$ and $d_1^{\rm odd} = -13\,000\pm 35\,000$ from BOSS, both of which are fully consistent with zero. For the mean of 2048 Patchy simulations, we find $d_0^{\rm odd} = -20\pm 770$ and $d_1^{\rm odd} = 0\pm 34\,000$, again indicating that the method is unbiased. Although the physical scale of the two coefficients is the same (in the Chern-Simons model they follow the relation $-d_0^{\rm odd} = d_1^{\rm odd}/3=A_{\rm CS}$), the first parameter is constrained almost $50\times$ better than the second: this is attributed to the different angular behavior of the two terms, with $d_0^{\rm odd}$ dominating the $\{\ell_1,\ell_2,\ell_3\}=\{1,1,1\}$ multiplet, for example (cf.\,Fig.\,\ref{fig: 4pcf-results-nseries}).

Our results for the Chern-Simons model may be compared to those obtained from the power spectrum and bispectrum of the CMB.\footnote{Since the coupling is assumed to be active only during inflation, the Lagrangian \eqref{eq: lag-cs} does not generate cosmic birefringence. Generation of parity-violating CMB spectra is possible however (due to helical gravitational wave production) but this is slow-roll suppressed, as noted above.} In particular, the \textit{Planck} 2018 $T$- and $E$-mode dataset (analyzed with the \textsc{smica} prescription) gave the constraints 
\beq
    -0.036 \leq g_\ast \leq 0.036, \quad -13 \leq c_0 \leq 11, \quad -7 \leq c_1 \leq 281, \quad -55 \leq c_2 \leq 37 
\eeq 
\citep{2020A&A...641A...9P,2020A&A...641A..10P}, on the inflationary parameters $g_\ast$ and $\{c_n\}$ appearing in the two- and three-point parametrizations of \eqref{eq: CS-2pcf} and \eqref{eq: CS-3pcf} at 95\% CL, and translating into our notation. Assuming $\gamma = 1$ and the above fiducial parameters, these can be used to place bounds on the gauge field energy density: $\rho_E/\rho_\phi\lesssim 7\times 10^{-13}$ ($2\times 10^{-16}$) using the two- (three-)point function measurements \citep[cf.][]{2016PhRvD..94h3503S}. Furthermore, a forecast of the detectability of $\rho_E/\rho_\phi$ from the CMB four-point function was presented in \citep{2016PhRvD..94h3503S}. This predicted a bound on $\rho_E/\rho_\phi$ of $\sim 3\times 10^{-20}$ at 95\% CL (or equivalently $\sigma\left(d_1^{\rm odd}\right)=640$) for a cosmic-variance dominated measurement using $\ell_{\rm max} = 2000$ and $|\gamma| = 1$. 

Bounds on the gauge field energy density from the BOSS 4PCF are far stronger than those obtained from the CMB anisotropic power spectrum and isotropic bispectrum (for $|\gamma|>0$), due to the exponential dependence on $|\gamma|$ \citep[Fig.\,4]{2016PhRvD..94h3503S}. As such, \textbf{they represent the strongest current constraints on Chern-Simons inflationary models.} Our measurement is roughly a factor of five worse than that predicted for the CMB: this occurs since the latter is contains significantly more Fourier-modes than the observed galaxy distribution, and thus an increased signal-to-noise ratio (although is subject to the smoothing effects of projection integrals). As the volume of spectroscopic data grows, we expect the constraints on $\rho_E/\rho_\phi$ to significantly strengthen, especially considering that the signal-to-noise of the Chern-Simons 4PCF \eqref{eq: zeta-cs-main-text} scales as $\sqrt{V_{\rm survey}}$, roughly independent of redshift.\footnote{Note that this differs from the signal-to-noise of the \textit{gravitational} 4PCF, which scales as $\left[b(z)D(z)\right]^2\sqrt{V_{\rm survey}}$ \citep{4pcf_boss}.} A survey such as DESI will probe $\sim 100\times$ the BOSS volume \citep{2016arXiv161100036D}, and should thus be expected to tighten the constraints on $\rho_E/\rho_\phi$ (and any other parity-breaking model amplitudes) by roughly an order of magnitude, providing stronger constraints on parity-breaking inflation than possible with the CMB. Finally, we note that, for LSS, the parity-odd 4PCF is an optimal place in which to search for these signatures, since, unlike other observables, the statistic is free from gravitational effects, thus we do not have to marginalize over the effects of late-time non-Gaussianity.


\section{Summary and Conclusions}\label{sec: summary}
Searching for parity-violation provides a unique manner in which to probe new physics occurring in the early Universe, including that of multi-field inflation, baryogenesis, and primordial magnetic field generation. Whilst there is a long history of constraining various parity-breaking phenomena using the CMB \citep{1999PhRvL..83.1506L,2008PhLB..660..444A,2012JCAP...06..015S,2015JCAP...07..039B,2021PhRvD.104b3507R,2016PhRvD..94h3503S,2017JCAP...07..034B,2011PhRvD..83b7301K,2011JCAP...06..003S}, few analyses have made use of LSS data. In this work, we have placed the first constraints on \unblind{(hitherto unexplored)} scalar-type parity-violation using the BOSS CMASS galaxy sample. The isotropic NPCFs are only parity-sensitive if $N>3$ \citep{2012PhRvL.108y1301J,2016PhRvD..94h3503S,2021arXiv211012004C}; recent developments in NPCF computation \citep{npcf_algo} have enabled efficient computation of the galaxy 4PCF, enabling such analyses. To provide a model-agnostic test, we have performed a blind search for parity-violation using the full parity-odd 4PCF (whose expectation value is zero in $\Lambda$CDM). Our primary tool has been a non-parametric rank test, comparing the BOSS 4PCF (on scales between $20\Mpch$ and $160\Mpch$) to that of a suite of realistic mock catalogs. This avoids the need to assume a Gaussian likelihood, and provides a robust (albeit conservative) \unblind{model-agnostic test. In the BOSS sample we found tentative evidence for parity-violation,} with a detection probability of \unblind{$99.6\%$, equivalent to $2.9\sigma$. \textbf{This indicates either new physics beyond the standard model or unknown systematics.}} 

As an additional test, we have performed a classical $\chi^2$-based analysis of the BOSS data, making use of a data compression scheme and a covariance matrix computed from mock catalogs. Furthermore, we use a theoretical Gaussian covariance \citep{npcf_cov} to facilitate high-fidelity compression, \unblind{which reduces the dataset to $N_{\rm eig}$ numbers}; importantly, the results are \textit{not} biased this choice, avoiding a potential systematic error. Post-compression, the empirical $\chi^2$ distribution closely matches that of the theory model \unblind{for $N_{\rm eig}\lesssim 100$; this gives credence to the assumption of Gaussianity.} For this test, we find a detection probability of \unblind{$83.3\%$} 
from BOSS \unblind{when using $N_{\rm eig} = 100$, or $100.0\%$ when using $N_{\rm eig}=250$, though the latter may be artificially inflated from likelihood non-Gaussianity. The results are broadly consistent with those from the rank test; however, the strong dependence on $N_{\rm eig}$ implies that our basis decomposition is inefficient, and that information may be being lost.}

We have a carried out a number of tests to explore potential systematic effects in our data which could lead to a false detection of parity violation. These include splitting the data into sub-regions, imposing radial and angular cuts, comparing against mock catalogs, altering the compression scheme, and normalizing by an overall rescaling factor. No clear evidence for systematics is observed, and we find our detection to be relatively coherent across various scales and sky regions. That said, our tests do rely heavily on the \textsc{Patchy} mocks well-representing the statistical properties of the BOSS data; although we have ruled out differences due to an overall rescaling factor, a scale dependent difference remains the most likely cause of our results, in the absence of a cosmological signal. 

Finally, we have used the measured 4PCF to bound the amplitudes of physical models of parity-violation. Here, we have primarily considered a single scenario; a Chern-Simons term in the inflationary Lagrangian, which couples the inflaton to a $U(1)$ gauge field. This leads to a definite prediction for the primordial polyspectra \citep{2015JCAP...07..039B,2016PhRvD..94h3503S}, which, with some effort, can be translated into a model for the galaxy 4PCF. Performing a Gaussian likelihood analysis using this template gives a comparable constraint on the ratio of gauge field and inflaton energy densities to that expected from the CMB \citep{2016PhRvD..94h3503S} (but much stronger than that from lower-order statistics); 
\unblind{$\rho_E/\rho_\phi<1.6\times 10^{-19}$} (95\% CL), 
assuming standard inflationary parameters. Notably, this does not appear to explain the above parity-violating signal. Similar constraints may be obtained for any other physical model giving a definite prediction for the galaxy 4PCF.

The coming years will lead to an explosion in the volume of LSS data available, \unblind{which will either confirm or refute the tentative detection of parity-violation found herein.}
Unlike for the gravitational contribution \citep{4pcf_boss}, the signal-to-noise of the inflationary 4PCF is not a strong function of redshift, with the constraints on models of new physics being primarily sensitive to the survey volume. To further increase the constraining power, we may fold in additional information, for example using the 5PCF and anisotropic NPCFs (which source additional information regarding vector parity-breaking \citep{2019arXiv190605198J}). Going beyond spectroscopic surveys, it is likely that high-volume observables such as intensity mapping and the Lyman-$\alpha$ forest, as well as the CMB itself, \unblind{will shed additional light on this}, 
pinning down a variety of new physics models.

\begin{acknowledgments}
\footnotesize
We thank Jiamin Hou for providing us with analytic covariance matrices. We are additionally indebted to Stephon Alexander, Giovanni Cabass, He Jia, Marc Kamionkowski, Morgane K\"onig and David Spergel for insightful discussions regarding parity-violation, Robert Cahn, Daniel Eisenstein, Jiamin Hou and Zachary Slepian for conversations relating to NPCFs, William Underwood for suggesting the non-parametric statistics of \S\ref{subsec: rank-test}, and Mikhail Ivanov, Marko Simonovi\'c, and Matias Zaldarriaga for general feedback. \resub{Furthermore, we are grateful to Paolo Campeti and Johannes Eskilt and the anonymous referee for providing valuable feedback on the manuscript.}
OHEP acknowledges funding from the WFIRST program through NNG26PJ30C and NNN12AA01C, thanks the Simons Foundation for additional support, \resub{and is grateful to the Institute for Advanced Study for their hospitality and abundance of baked goods}.

\vskip 2pt

The authors are pleased to acknowledge that the work reported in this paper was substantially performed using the Princeton Research Computing resources at Princeton University, which is a consortium of groups led by the Princeton Institute for Computational Science and Engineering (PICSciE) and the Office of Information Technology's Research Computing Division.

Funding for SDSS-III has been provided by the Alfred P. Sloan Foundation, the Participating Institutions, the National Science Foundation, and the U.S. Department of Energy Office of Science. The SDSS-III web site is \href{http://www.sdss3.org/}{www.sdss3.org}.

SDSS-III is managed by the Astrophysical Research Consortium for the Participating Institutions of the SDSS-III Collaboration including the University of Arizona, the Brazilian Participation Group, Brookhaven National Laboratory, Carnegie Mellon University, University of Florida, the French Participation Group, the German Participation Group, Harvard University, the Instituto de Astrofisica de Canarias, the Michigan State/Notre Dame/JINA Participation Group, Johns Hopkins University, Lawrence Berkeley National Laboratory, Max Planck Institute for Astrophysics, Max Planck Institute for Extraterrestrial Physics, New Mexico State University, New York University, Ohio State University, Pennsylvania State University, University of Portsmouth, Princeton University, the Spanish Participation Group, University of Tokyo, University of Utah, Vanderbilt University, University of Virginia, University of Washington, and Yale University. 

The massive production of all MultiDark-Patchy mocks for the BOSS Final Data Release has been performed at the BSC Marenostrum supercomputer, the Hydra cluster at the Instituto de Fısica Teorica UAM/CSIC, and NERSC at the Lawrence Berkeley National Laboratory. We acknowledge support from the Spanish MICINNs Consolider-Ingenio 2010 Programme under grant MultiDark CSD2009-00064, MINECO Centro de Excelencia Severo Ochoa Programme under grant SEV- 2012-0249, and grant AYA2014-60641-C2-1-P. The MultiDark-Patchy mocks was an effort led from the IFT UAM-CSIC by F. Prada’s group (C.-H. Chuang, S. Rodriguez-Torres and C. Scoccola) in collaboration with C. Zhao (Tsinghua U.), F.-S. Kitaura (AIP), A. Klypin (NMSU), G. Yepes (UAM), and the BOSS galaxy clustering working group.
\end{acknowledgments}

\appendix 

\section{Likelihood Non-Gaussianity}\label{appen: NG-likelihood}
A major assumption of most cosmological analyses is that the underlying likelihood for the statistic of interest can be well approximated as Gaussian. Whilst this is often ensured by the central limit theorem, it can break down in the case of highly-correlated data, such as that considered in this work. In this appendix, we present a simple test to check whether the likelihood of the full 4PCF can be justifiably considered Gaussian.

For this purpose, we take the \textsc{Patchy-Nseries} NGC simulations (\S\ref{subsec: nseries-testing}) and partition them into two groups. The first 500 are used to compute a sample mean and covariance for the distribution, thus defining a Gaussian distribution from which we draw $10^5$ mock observations. Each of these is compressed into one dimension via the \textit{pseudo}-$\chi^2$ statistic defined in \eqref{eq: pseudo-chi2}, and histogrammed. This can then be compared to the empirical \textit{pseudo}-$\chi^2$ distribution obtained from the remaining simulations directly. If the likelihood is Gaussian, the two distributions should match.

\begin{figure}
    \centering
    \includegraphics[width=0.5\textwidth]{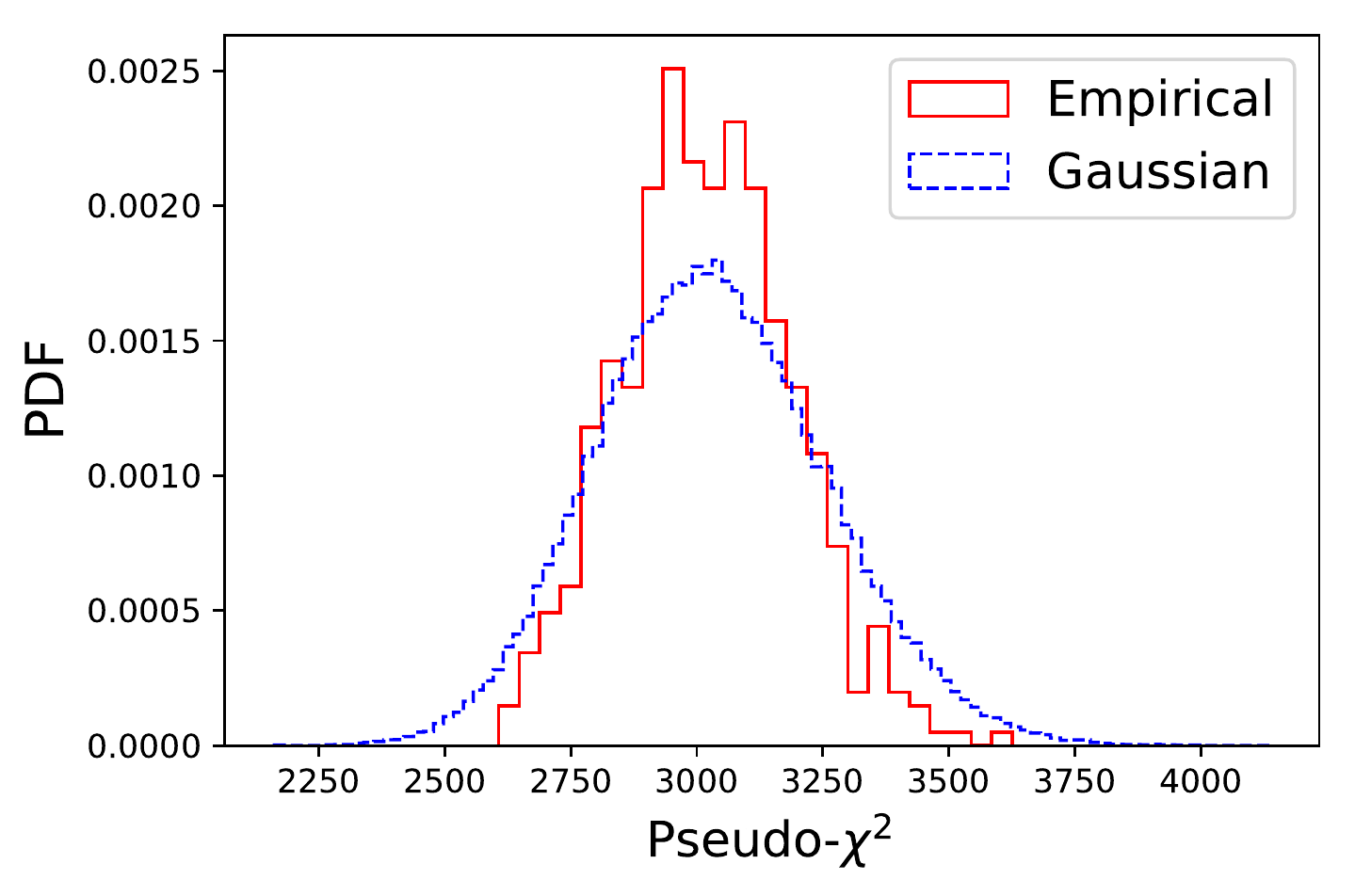}
    \caption{Distribution of the \textit{pseudo}-$\chi^2$ statistic \eqref{eq: pseudo-chi2} estimated from \textsc{Patchy-Nseries} simulations (red) and Gaussian realizations (blue), drawn from a distribution with mean and covariance estimated from a second set of \textsc{Patchy-Nseries} simulations. The clear differences between the distributions indicates that the full 4PCF likelihood cannot be well-approximated as Gaussian.}
    \label{fig: likNG}
\end{figure}

The resulting PDFs are shown in Fig.\,\ref{fig: likNG}. Notably, the empirical and Gaussianized distributions do not match, indicating that the full 4PCF distribution is not well described by a Gaussian, even in the best-case scenario when the mean and variance are estimated from the simulations. In particular, the Gaussian assumption overestimates the sample variance, which will reduce any potential detection significance. If one instead uses the \textit{theoretical} covariance to define the Gaussian distribution, the situation is far worse: the \textit{pseudo}-$\chi^2$ distribution instead peaks at $\approx 1250$, indicating a breakdown of the modelling assumptions (as discussed in \S\ref{subsec: analytic-cov}). Finally, we note that, although the full (uncompressed) 4PCF distribution appears to be non-Gaussian, this does not imply that the same is true for the projected statistics of \S\ref{subsec: compressed-analysis}, since the central limit theorem becomes more applicable as the dimensionality reduces.

\section{Derivation of the Chern-Simons 4PCF Model}\label{appen: chern-simons}
Below, we sketch the derivation of the parity-odd 4PCF induced by the Chern-Simons coupling of \S\ref{sec: inflationary-model}. Our starting point is the general expression given in \eqref{eq: CS-4PCF-overall}, which is a product of four pieces. By expanding the angular dependence of each term using the isotropic basis functions (\S\ref{subsec: basis}), we may compute the full 4PCF efficiently.

To begin, we consider the primordial Chern-Simons trispectrum defined in \eqref{eq: tk-CS}. The angular pieces may be written in terms of isotropic basis functions of three coordinates using \citep[Appendix A.2]{2020arXiv201014418C}:
\beq
    (\hk_1\times\hk_3)\cdot\hs = i\frac{\sqrt{2}}{3}(4\pi)^{3/2}\P_{111}(\hk_1,\hk_3,\hs),\qquad \hk_1\cdot\hk_3 = -\frac{1}{\sqrt{3}}(4\pi)^{3/2}\P_{110}(\hk_1,\hk_3,\hs),\\\nonumber
    \hk_1\cdot\hs = -\frac{1}{\sqrt{3}}(4\pi)^{3/2}\P_{101}(\hk_1,\hk_3,\hs),\qquad 
    \hk_1\cdot\hk_3 = -\frac{1}{\sqrt{3}}(4\pi)^{3/2}\P_{011}(\hk_1,\hk_3,\hs).
\eeq
The resulting products of two basis functions can be simplified using \citep[\S6.3]{2020arXiv201014418C}, yielding
\beq
    t(\vk_1,\vk_3,\vs) &=& -\sqrt{2}A_{\rm CS}(4\pi)^{3/2}P_\zeta(k_1)P_\zeta(k_3)P_\zeta(s)\\\nonumber
    &&\,\times\,\left[\P_{111}(\hk_1,\hk_3,\hs)+\frac{1}{\sqrt{5}}\P_{221}(\hk_1,\hk_3,\hs)+\frac{1}{\sqrt{5}}\P_{212}(\hk_1,\hk_3,\hs)-\frac{1}{\sqrt{5}}\P_{122}(\hk_1,\hk_3,\hs)\right],
\eeq
whose radial part is separable in $k_1$, $k_3$ and $s$. For the general parity-odd trispectrum given in \eqref{eq: tk-even-odd}, we obtain the same result, but with the replacement $A_{\rm CS}\P_{111}\to -d_0^{\rm odd}\P_{111}$, $A_{\rm CS}\P_{221}+\text{2 perms.}\to d_1^{\rm odd}\P_{221}/3+\text{2 perms.}$.

Next, we consider the Dirac delta functions. By rewriting $\delD{\vk_1+\vk_2-\vs}$ as a complex exponential, inserting plane-wave expansions \citep[Eq.\,16.63]{arfken2013mathematical}, then performing the angular integral, the function can be expressed as a sum over one-dimensional integrals and isotropic basis functions of three coordinates:
\beq
    \delD{\vk_1+\vk_2-\vs} &=& (4\pi)^{5/2}\sum_{L_1L_2L_5}i^{L_1+L_2+L_5}(-1)^{L_5}\mathcal{C}_{L_1L_2L_5}\tjo{L_1}{L_2}{L_5}\P_{L_1L_2L_5}(\hk_1,\hk_2,\hs)\\\nonumber
    &&\qquad\,\times\, \int_0^\infty x^2dx\,j_{L_1}(k_1x)j_{L_2}(k_2x)j_{L_5}(sx),
\eeq
where $\mathcal{C}_{L_1\ldots L_N} \equiv \sqrt{(2L_1+1)\ldots(2L_N+1)}$, the $2\times 3$ matrices are Wigner 3-$j$ symbols and we have used properties of the Gaunt integral \citep[Eq.\,34.3.22]{nist_dlmf}. Similarly,
\beq
    \delD{\vk_3+\vk_4+\vs} &=& (4\pi)^{5/2}\sum_{L_3L_4L_5'}i^{L_3+L_4+L_5'}\mathcal{C}_{L_3L_4L_5'}\tjo{L_3}{L_4}{L_5'}\P_{L_3L_4L_5'}(\hk_3,\hk_4,\hs)\\\nonumber
    &&\,\qquad\times\,\int_0^\infty x'^2dx'\,j_{L_3}(k_3x')j_{L_4}(k_4x')j_{L_5'}(sx').
\eeq
Note that the integrands are again separable in $\{k_i\}$ and $s$. Using the approach of \citep{2021RSPSA.47710376P}, they may equivalently be rewritten as infinite sums.

For the transfer functions $M(k,z)Z_1(\vk,z)$, we first expand the redshift-space kernel $Z_1(\vk)$ in spherical harmonics (dropping the redshift dependence for clarity):
\beq
    Z_1(\hk;\hn) \equiv b + f(\hk\cdot\hn)^2 = 4\pi\sum_{\ell m}\left[\left(b+\frac{f}{3}\right)\delta^{\rm K}_{\ell 0}+\frac{2f}{15}\delta^{\rm K}_{\ell 2}\right]Y_{\ell m}^*(\hn)Y_{\ell m}(\hk) \equiv 4\pi\sum_{\ell m}Z_\ell\,Y^*_{\ell m}(\hn)Y_{\ell m}(\hk),
\eeq
for linear bias $b(z)$, growth rate $f(z)$ and line-of-sight $\hn$. Since we consider only isotropic 4PCFs in this work, we can integrate over the LoS orientation (which is equivalent to performing a joint rotation of all $\{\vr_i\}$). Following some algebra, this leads to a set of isotropic functions of \textit{four} coordinates (see \citep{2020arXiv201014418C} for details):
\beq
    \int\frac{d\hn}{4\pi}Z_1(\hk_1;\hn)Z_1(\hk_2;\hn)Z_1(\hk_3;\hn)Z_1(\hk_4;\hn) &=& (4\pi)^2\sum_{j_1j_2j_{12}j_3j_4}\tjo{j_1}{j_2}{j_{12}}\tjo{j_{12}}{j_3}{j_4}Z_{j_1}Z_{j_2}Z_{j_3}Z_{j_4}\\\nonumber
    &&\,\times\,\mathcal{C}_{j_1j_2j_{12}j_3j_4}\P_{j_1j_2(j_{12})j_3j_4}(\hk_1,\hk_2,\hk_4,\hk_4),
\eeq
where $j_i\in\{0,2\}$ and $j_{12}\in\{0,2,4\}$.

The final contribution is from the Fourier basis functions and their permutations, which can be written
\beq
    e^{i(\vk_1\cdot\vr_1+\vk_2\cdot\vr_2+\vk_3\cdot\vr_3)}+\text{23 perms.} &=& \sum_H e^{i(\vk_1\cdot\vr_{H1}+\vk_2\cdot\vr_{H2}+\vk_3\cdot\vr_{H3}+\vk_4\cdot\vr_{H4})},
\eeq
where $\{H1,H2,H3,H4\}$ is one of the 24 permutations of $\{1,2,3,4\}$, and we have introduced $\vr_4 = 0$ for convenience. Projecting onto the 4PCF basis functions $\P_{\ell_1\ell_2\ell_3}(\hr_1,\hr_2,\hr_3)$ gives a sum of isotropic functions of four coordinates:
\beq
    &&\sum_H(4\pi)^{-1/2}\int d\hr_1d\hr_2d\hr_3d\hr_4\,\P^*_{\ell_1\ell_2\ell_3}(\hr_1,\hr_2,\hr_3)e^{i(\vk_1\cdot\vr_{H1}+\vk_2\cdot\vr_{H2}+\vk_3\cdot\vr_{H3}+\vk_4\cdot\vr_{H4})}Y_{\ell_4m_4}(\hr_4)\\\nonumber
    &&=\sum_H(4\pi)^{7/2}\Phi_H(-i)^{\ell_1+\ell_2+\ell_3}j_{\ell_{H1}}(k_1r_{H1})j_{\ell_{H2}}(k_2r_{H2})j_{\ell_{H3}}(k_3r_{H3})j_{\ell_{H4}}(k_4r_{H4})\P_{\ell_{H1}\ell_{H2}(\ell^*)\ell_{H3}\ell_{H4}}(\hk_1,\hk_2,\hk_3,\hk_4),
\eeq 
using the plane wave expansion and inserting $\ell_4=m_4=r_4=0$ in the first line. In the second line we include a permutation factor $\Phi_H$, given by $(-1)^{\ell_1+\ell_2+\ell_3}$ if $\{\ell_{H1},\ell_{H2},\ell_{H3},\ell_{H4}\}$ is an odd permutation of $\{\ell_1,\ell_2,\ell_3\}$ (removing the zero element) and unity else. Furthermore, $\ell^*$ is set by the position of the zero, e.g., $\ell^* = \ell_{H2}$ if $\ell_{H1} = 0$, $\ell^* = \ell_{H4}$ if $\ell_{H3}=0$ \textit{et cetera}.

Combining the above results, we obtain
\beq\label{eq: zeta-cs-appendix}
    \zeta_{\ell_1\ell_2\ell_3}(r_1,r_2,r_3) &=& (4\pi)^{11}\sqrt{2}A_{\rm CS}\,i^{\ell_1+\ell_2+\ell_3}\sum_H\sum_{L_1L_2L_3L_4L_5L_5'}i^{L_1+L_2+L_3+L_4-L_5+L_5'}\mathcal{C}_{L_1L_2L_3L_4L_5L_5'}\\\nonumber
    &&\,\times\,\tjo{L_1}{L_2}{L_5}\tjo{L_3}{L_4}{L_5'}\times\,\mathcal{M}^{\ell_{H1}\ell_{H2}(\ell^*)\ell_{H3}\ell_{H4}}_{L_1L_2L_3L_4L_5L_5'}\\\nonumber
    &&\,\times\,\int x^2dx\,\int x'^2dx'\,K_{L_5L_5'}(x,x')I_{L_1}^{\ell_{H1}}(x;r_{H1})J_{L_2}^{\ell_{H2}}(x;r_{H2})I_{L_3}^{\ell_{H3}}(x';r_{H3})J_{L_4}^{\ell_{H4}}(x';r_{H4}),
\eeq
defining the integrals:
\beq\label{eq: I,J,K-def}
    &&I_{L}^{\ell}(x;r) \equiv \int_0^\infty\frac{k^2dk}{2\pi^2}M(k)P_\zeta(k)j_L(kx)j_\ell(kr), \qquad J_L^{\ell}(x;r) \equiv \int_0^\infty\frac{k^2dk}{2\pi^2}M(k)j_L(kx)j_\ell(kr),\\\nonumber
    &&\qquad\qquad\qquad\qquad K_{LL'}(x,x') \equiv \int_0^\infty\frac{s^2ds}{2\pi^2}P_\zeta(s)j_{L}(sx)j_{L'}(sx').
\eeq
In practice, we must integrate the statistic over radial bins of finite width, which corresponds to replacing e.g., $j_{\ell}(kr)$ with $\bar\jmath_{\ell}^{b}(k)$ for bin $b$. The bin-integrated Bessel functions are analytic and their forms can be found in \citep[Appendix A]{2020MNRAS.492.1214P}. 

The coupling matrix in \eqref{eq: zeta-cs-appendix} is given by an integral over five isotropic basis functions of five coordinates:
\beq\label{eq: coupling-matrix}
    {\mathcal M}^{\ell_{H1}\ell_{H2}(\ell^*)\ell_{H3}\ell_{H4}}_{L_1L_2L_3L_4L_5L_5'} = &&\,\sum_{j_1j_2j_{12}j_3j_4}\mathcal{C}_{j_1j_2j_{12}j_3j_4}\tjo{j_1}{j_2}{j_{12}}\tjo{j_{12}}{j_3}{j_4}Z_{j_1}Z_{j_2}Z_{j_3}Z_{j_4}\int d\hk_1d\hk_2d\hk_3d\hk_4d\hs\\\nonumber
    &&\,\times\,\left[\P_{j_1j_2(j_{12})j_3(j_4)j_40}\P_{L_1L_2(L_5)0(L_5)0L_5}\P_{00(0)L_3(L_3)L_4L_5'}\P_{\ell_{H1}\ell_{H2}(\ell^*)\ell_{H3}(\ell_{H4})\ell_{H4}0}\right](\hk_1,\hk_2,\hk_3,\hk_4,\hs)\\\nonumber
    &&\,\times\,\left[\P_{10(1)1(1)01}+\frac{1}{\sqrt{5}}\P_{20(2)2(2)01}+\frac{1}{\sqrt{5}}\P_{20(2)1(2)02}-\frac{1}{\sqrt{5}}\P_{10(1)2(2)02}\right](\hk_1,\hk_2,\hk_3,\hk_4,\hs),
\eeq
where we have noted that that isotropic functions of $N$ coordinates may be rewritten in terms of those with $N+M\geq N$ coordinates by inserting a factor $(4\pi)^{M/2}$. Despite its complexity, this can be evaluated analytically, making extensive use of the product relation for isotropic basis functions of five coordinates:
\beq
    \P_\L\P_{\L'} &=& (4\pi)^{-5/2}\sum_{\L''}(-1)^{\L_1''+\L_2''+\L_3''+\L_4''+\L_5''}\mathcal{C}_{\L}\mathcal{C}_{\L'}\mathcal{C}_{\L''}\prod_{i=1}^5\left[\tjo{\L_i}{\L'_i}{\L_i''}\right]\P_{\L''}\\\nonumber
    &&\,\times\,\begin{Bmatrix}\L_1&\L_2&\L_{12}\\\L_1'&\L_2'&\L_{12}'\\\L_1''&\L_2''&\L_{12}''\end{Bmatrix}\begin{Bmatrix}\L_{12}&\L_3&\L_{123}\\\L_{12}'&\L_3'&\L_{123}'\\\L_{12}''&\L_3''&\L_{123}''\end{Bmatrix}\begin{Bmatrix}\L_{123}&\L_4&\L_{5}\\\L_{123}'&\L_{4}'&\L_{5}'\\\L_{123}''&\L_4''&\L_{5}''\end{Bmatrix}
\eeq
\citep[\S6.5]{2020arXiv201014418C}, where $\L\equiv\{\L_1,\L_2,(\L_3),\L_4,(\L_{123}),\L_4,\L_5\}$, the curly braces indicate Wigner 9-$j$ symbols and $\mathcal{C}_{\L}$ involves all elements of $\L$. This simplifies considerably when some elements of $\L$ or $\L'$ are zero. 

\bibliographystyle{JHEP}
\bibliography{adslib,otherlib}

\end{document}